\documentclass[twocolumn, tighten]{aastex63}

\usepackage{amsmath,amsthm,amssymb}
\usepackage{multirow}
\usepackage{hyperref}
\usepackage{xspace}
\usepackage{epstopdf}
\usepackage{graphicx}

\newcommand\superbit{\textsc{SuperBIT}\xspace}
\newcommand\uoftastro{\affiliation{David A. Dunlap Dept. of Astronomy and Astrophysics, University of Toronto, 50 St. George Street, Toronto, ON, Canada M5S 3H4}}
\newcommand\dunlap{\affiliation{Dunlap Institute for Astronomy and Astrophysics, University of Toronto, 50 St. George Street, Toronto, ON, Canada M5S 3H4}}
\newcommand\uoftphysics{\affiliation{Department of Physics, University of Toronto, 60 St. George Street, Toronto, ON, Canada M5R 2M8}}
\newcommand\utias{\affiliation{University of Toronto Institute for Aerospace Studies (UTIAS), 4925 Dufferin Street, Toronto, ON, Canada M3H 5T6}}
\newcommand\princeton{\affiliation{Department of Physics, Princeton University, Jadwin Hall, Princeton, NJ, USA 08544}}
\newcommand\durhamcfai{\affiliation{Centre for Advanced Instrumentation (CfAI), Durham University, South Road, Durham DH1 3LE, UK}}
\newcommand\stewardobs{\affiliation{Department of Astronomy/Steward Observatory, University of Arizona, 933 North Cherry Avenue, Tucson, AZ, USA 85721}}
\newcommand\jpl{\affiliation{Jet Propulsion Laboratory (JPL), California Institute of Technology, 4800 Oak Grove Drive, Pasadena, CA, USA 91109}}
\newcommand\durhamcfea{\affiliation{Centre for Extragalactic Astronomy, Department of Physics, Durham University, Durham DH1 3LE, UK}}
\newcommand\durhamifcc{\affiliation{Institute for Computational Cosmology, Durham University, South Road, Durham DH1 3LE, UK}}
\newcommand\oslo{\affiliation{Institute of Theoretical Astrophysics, University of Oslo, Blindern, Oslo 0315, Norway}}
\newcommand\sheffield{\affiliation{Department of Physics and Astronomy, The University of Sheffield, Hounsfield Road, Sheffield S3 7RH, UK}}
\newcommand\washu{\affiliation{Department of Physics, Washington University in St. Louis, 1 Brookings Drive, St. Louis, MO, USA, 63130}}
\newcommand\mcdonnell{\affiliation{McDonnell Center for the Space Sciences, Washington University in St. Louis, 1 Brookings Dr., St. Louis, MO USA 63130}}

\begin{document}

\title{Optical night sky brightness measurements from the stratosphere}

\correspondingauthor{Ajay Gill}
\email{ajay.gill@mail.utoronto.ca}

\author[0000-0002-3937-4662]{Ajay Gill}
\uoftastro{}
\dunlap{}

\author{Steven J. Benton}
\princeton{}

\author{Anthony M.\ Brown}
\durhamcfai
\durhamcfea

\author{Paul Clark}
\durhamcfai

\author{Christopher J.\ Damaren}
\utias

\author{Tim Eifler}
\stewardobs

\author{Aurelien A.\ Fraisse}
\princeton

\author{Mathew N. Galloway}
\oslo

\author{John W.\ Hartley}
\uoftphysics

\author{Bradley Holder}
\utias
\dunlap

\author{Eric M.\ Huff}
\jpl

\author{Mathilde Jauzac}
\durhamcfea
\durhamifcc

\author{William C.\ Jones}
\princeton

\author{David Lagattuta}
\durhamcfea

\author{Jason S.-Y. Leung}
\uoftastro
\dunlap

\author{Lun Li}
\princeton

\author{Thuy Vy T.\ Luu}
\princeton

\author{Richard J.\ Massey}
\durhamcfai
\durhamcfea
\durhamifcc

\author{Jacqueline McCleary}
\jpl

\author{James Mullaney}
\sheffield

\author{Johanna M. Nagy}
\dunlap
\washu
\mcdonnell

\author{C.\ Barth Netterfield}
\uoftastro
\dunlap
\uoftphysics

\author{Susan Redmond}
\princeton

\author{Jason D.\ Rhodes}
\jpl

\author{L. Javier Romualdez}
\princeton

\author{J\"urgen Schmoll}
\durhamcfai

\author{Mohamed M.\ Shaaban}
\dunlap
\uoftphysics

\author{Ellen Sirks}
\durhamifcc

\author{Suresh Sivanandam}
\uoftastro
\dunlap

\author{Sut-Ieng Tam}
\durhamcfea

\begin{abstract}
This paper presents optical night sky brightness measurements from the stratosphere using CCD images taken with the Super-pressure Balloon-borne Imaging Telescope (\superbit). The data used for estimating the backgrounds were obtained during three commissioning flights in 2016, 2018, and 2019 at altitudes ranging from 28 km to 34 km above sea level. For a valid comparison of the brightness measurements from the stratosphere with measurements from mountain-top ground-based observatories (taken at zenith on the darkest moonless night at high Galactic and high ecliptic latitudes), the stratospheric brightness levels were zodiacal light and diffuse Galactic light subtracted, and the airglow brightness was projected to zenith. The stratospheric brightness was measured around 5.5 hours, 3 hours, and 2 hours before the local sunrise time in 2016, 2018, and 2019 respectively. The $B$, $V$, $R$, and $I$ brightness levels in 2016 were 2.7, 1.0, 1.1, and 0.6 mag arcsec$^{-2}$ darker than the darkest ground-based measurements. The $B$, $V$, and $R$ brightness levels in 2018 were 1.3, 1.0, and 1.3 mag arcsec$^{-2}$ darker than the darkest ground-based measurements.  The $U$ and $I$ brightness levels in 2019 were 0.1 mag arcsec$^{-2}$ brighter than the darkest ground-based measurements, whereas the $B$ and $V$ brightness levels were 0.8 and 0.6 mag arcsec$^{-2}$ darker than the darkest ground-based measurements. The lower sky brightness levels, stable photometry, and lower atmospheric absorption make stratospheric observations from a balloon-borne platform a unique tool for astronomy. We plan to continue this work in a future mid-latitude long duration balloon flight with \superbit{}.
\end{abstract}


\section{Introduction} \label{sec:intro}

The objective when doing photometry is to determine the true brightness of the individual astronomical source of interest. However, various sources of sky brightness can contaminate the flux from astronomical sources. For the case of aperture photometry, the signal-to-noise ratio of a measurement is given by the equation \citep{mortara}

\begin{equation}
    \frac{S}{N} = \frac{N_{*}}{\sqrt{N_{*} + n_{\rm pix}(N_{\rm S} + N_{\rm D} + N_{\rm R}^{2})}}
\end{equation}

\noindent where $N_{*}$ is the number of photons collected from the source of interest (or the ``signal"). The ``noise" terms in the equation are the square roots of $N_{*}$ plus $n_{\rm pix}$ (the number of pixels under consideration for the $S/N$ calculation) times the contributions from $N_{\rm S}$ (the total number of photons per pixel from the background or the sky), $N_{\rm D}$ (the total number of dark current electrons per pixel), and $N_{\rm R}^{2}$ (the total number of electrons per pixel from read noise). For observations in the sky background limited case, such that $\sqrt{n_{\rm pix}N_{\rm S}} > 3 \sqrt{n_{\rm pix}N_{\rm R}^{2}}$, the $S/N$ is approximately

\begin{equation}
    \frac{S}{N} \simeq \frac{N_{*}}{\sqrt{n_{\rm pix} N_{\rm S}}} 
\end{equation}

Understanding the sky background level at the observing site is therefore important, as it can set the limiting magnitude for detection of astronomical sources. There are a variety of sources of different physical origin that can contribute to the total night sky background. We refer the reader to \citet{roach_gordon_1973} and \citet{leinert} for a comprehensive review. 

\textit{Zodiacal light} ($I_{\rm ZL}$) in the UV, visual, and near-IR is caused by sunlight scattered from the diffuse cloud of interplanetary dust particles that lies primarily in the plane of the solar system. In the mid- and far-IR, $I_{\rm ZL}$ is dominated by the thermal emission from those dust particles. $I_{\rm ZL}$ is a function of the viewing direction ($\lambda - \lambda_{\odot}$, $\beta$), wavelength, heliocentric distance, and the position of the observer relative to the symmetry plane of interplanetary dust. $I_{\rm ZL}$ is also polarized, with a maximum polarization of $\sim$ 20$\%$ \citep{leinert}. $I_{\rm ZL}$ as a function of ecliptic coordinates in the optical wavelengths has been measured both from the ground and from space by a few different studies (see e.g. \citealp{KWON200491, BUFFINGTON201688, Lasue_2020}). 

\textit{Airglow} ($I_{\rm A}$) due to the chemiluminescence of upper atmosphere atoms and molecules can also contribute to night sky brightness and is a function of zenith angle, local time, geographic latitude, season, solar activity, and altitude. Airglow includes a quasi-continuum from NO$_{2}$ (500 - 650 nm) and a number of discrete emission lines. Airglow emission lines mainly arise from the thin mesospheric layer at an altitude of $\sim$ 85 to 90 km (see e.g. \citealp{meinel_one, meinel_two, chamberlain, roach_1964, roach_gordon_1973, meier_1991, keyton}). The strongest airglow line in the visible is the 557.7 nm forbidden line of [OI]. OH lines dominate the airglow emission in the near-IR bands \citep{meinel_one, meinel_two}. We refer the reader to Table 13 in \citet{leinert} for a list of airglow lines along with their emission wavelength, typical altitude of the atmospheric emission layer, and typical intensities. In the absence of atmospheric extinction, a thin homogeneous emiting layer at height $h$ above the Earth's surface shows an increase in airglow brightness towards the horizon described by the \textit{van Rhijn function} \citep{vanRhijn}

\begin{equation}
    \frac{I(z)}{I(\text{zenith})} = \frac{1}{\sqrt{1 - [R / (R + h)]^{2} \sin^{2}z}}
    \label{vanRhijn_func}
\end{equation}

\noindent where $R$ = 6378 km is the radius of the Earth and $z$ is the zenith distance. The increase in airglow brightness towards the horizon has been observationally verified to be consistent with the van Rhijn function (see e.g. \citealp{Hofmann_77} for measurements taken with balloon observations at 2.1 $\mu$m at an altitude of 30 km). 

\textit{Integrated starlight} ($I_{\rm ISL}$) is the combined light from unresolved stars in the Milky Way that contribute to the sky brightness from the UV to mid-IR, with the contribution dominated by hot stars and white dwarfs at the shortest wavelengths, main sequence stars in the visible, and red giants in the IR \citep{mathis}. The contribution of $I_{\rm ISL}$ depends on the ability for the telescope to resolve the brightness stars, which is set by its limiting magnitude. The limiting magnitude of a telescope depends on the \textit{seeing} at the site, the atmospheric extinction, and the size of the telescope. 

\textit{Diffuse Galactic light} ($I_{\rm DGL}$) is due to the diffuse component of the Galactic background radiation produced by scattering of starlight by interstellar dust  \citep{elvey1937, roach_gordon_1973}. The scattering of starlight by interstellar dust is the primary contributor to the interstellar extinction of starlight. Therefore, $I_{\rm DGL}$ is brightest in directions where both the dust column density and the integrated stellar emissivity are high, which is generally the case for the lowest Galactic latitudes. $I_{\rm DGL}$ typically contributes $\sim 20 - 30 \%$ of the total integrated light from the Milky Way \citep{leinert}. $I_{\rm DGL}$ is difficult to measure from ground-based observations, since the contribution from $I_{A}$, $I_{\rm ZL}$, and $I_{\rm ISL}$ must all be known to very high precision if the $I_{\rm DGL}$ component is to estimated by subtraction of the other components. 

\textit{Extragalactic background light} ($I_{\rm EBL}$) due to redshifted starlight from unresolved galaxies, stars or gas in intergalactic space, or redshifted emission from dust particles heated by starlight in galaxies can also contribute to the total sky background. Although no generally acceptable measurements exist in the UV, optical, or IR wavebands, the contribution of $I_{\rm EBL}$ is expected to be very small at all sites. 

Small imaging photopolarimeters (IPP's) on the Pioneer 10 and 11 deep space probes were used during the cruise phases (between and beyond the planets) to periodically measure and map the sky brightness and polarization in blue (395 nm - 495 nm) and red (590 nm - 690 nm) bands from beyond the asteroid belt ($R > $ 3 AU), where the contribution of zodiacal light is negligible \citep{weinberg_1974, hanner}. \citet{toller_1981} derived $I_{\rm DGL}$ intensities in the blue band from the Pioneer 10 data by subtracting the $I_{\rm ISL}$ measured by \citet{megill} and \citet{sharov} at the positions of 194 Selected Areas \citep{blauuw}. The residuals are interpreted to be largely due to the contribution of $I_{\rm DGL}$. Figure 76 in \citet{leinert} presents the mean Galactic latitude dependence of $I_{\rm DGL}$ from \citet{toller_1981}, averaged over all Galactic longitudes.

\textit{Moonlight} ($I_{\rm Moon}$) can also contribute to sky brightness and is a function of lunar phase and the moon-target angular separation. \citet{Krisciunas_1991} provide a model for the sky brightness due to moonlight as a function of the moon's phase, the zenith distance of the moon, the zenith distance of the sky position, the angular separation of the moon and sky position, and the local extinction coefficient. \citet{Jones_2013} developed an advanced scattered moonlight model for Cerro Paranal, which can be modified for any location with known atmospheric properties. \citet{walker} also found correlation between solar activity and the $V$ and $B$-band zenith sky brightness using photometric measurements at the San Benito Mountain (1.6 km above sea level) during 1976 to 1987. 

The combined radiation from the different components of sky brightness is attenuated by \textit{atmospheric extinction}, while \textit{tropospheric scattering} ($I_{\rm sca}$) of the incoming radiation also adds a non-negligible brightness component. $I_{\rm sca}$ also contains a contribution from light pollution. The total sky background can be expressed as

\begin{equation}
    I_{\rm{sky}} = (I_{\rm A} + I_{\rm ZL} + I_{\rm ISL} + I_{\rm DGL} + I_{\rm EBL} + I_{\rm Moon})\,\cdot\, \mathrm{e}^{-\tau} + I_{\rm sca}  
\end{equation}

\noindent where $\tau$ is the \textit{extinction coefficient} (which depends on the wavelength $\lambda$, zenith distance $z$, height of the observer, and the change of the atmospheric conditions with time). For observations from the stratosphere, the atmospheric extinction is negligible and tropospheric scattering is irrelevant, such the total sky background from stratospheric altitudes can be approximated as

\begin{equation}
    I_{\rm{sky}} \simeq I_{\rm A} + I_{\rm ZL} + I_{\rm ISL} + I_{\rm DGL} + I_{\rm EBL} + I_{\rm Moon}
\end{equation}

There have been a number of studies that have estimated the optical sky background from ground-based observatories. \citet{Benn_1998} estimated the brightness on the island of La Palma in the Canary Islands using 427 CCD images taken with the Isaac Newton and Jacobus Kapteyn Telescopes on 63 nights from 1987 to 1996. These telescopes are located at longitude $18^{\circ}$ W, latitude $20^{\circ}$ N, and an altitude of 2.3 km above sea level. Their zenith sky brightness measurements on moonless nights at high ecliptic and Galactic latitudes, low airmass, and at solar minimum are 22.0, 22.7, 21.9, 21.0, and 20.0 mag arcsec$^{-2}$ in $U, B, V, R$ and $I$ respectively. Optical sky brightness has been measured from the Gemini North Observatory\footnote{https://www.gemini.edu/observing/telescopes-and-sites/sites\#OptSky} (located near the summit of Mauna Kea at 4.2 km above sea level). \citet{Krisciunas_1997} measured the average zenith sky brightness levels during moonless nights at the 2.8 km level at Mauna Kea between 1985 to 1996 to be 22.5 and 21.6 mag arcsec$^{-2}$ in $B$ and $V$ respectively.

\citet{leinert_1995} present sky brightness measurements taken from the Calar Alto Observatory at an altitude of 2.17 km during 18 moonless nights in the years 1989, 1990, 1991, and 1993. Their average values are 22.2, 22.6, 21.5, 20.6, and 18.7 mag arcsec$^{-2}$ in $UBVRI$. They also found that long-term variations in sky brightness are correlated with the solar activity. \citet{mattila} measured the sky brightness at the La Silla Observatory located at an altitude of 2.4 km during 40 moonless nights between 1978 to 1988. They found their results to be 22.8, 21.7, 20.8, and 19.5 mag arcsec$^{-2}$ in $B, V, R$ and $I$ respectively. 

\citet{patat} measured the optical sky brightness at the Paranal Observatory (2.64 km above sea level) using 3900 images obtained on 174 different nights from April 2000 and September 2001. Their zenith-corrected values averaged over the whole period are 22.3, 22.6, 21.6, 20.9, and 19.7 mag arcsec$^{-2}$ in $U, B, V, R$ and $I$ respectively. \citet{Yang_2017} measured the optical sky brightness at the summit of the Antarctic plateau, Dome A (located 4.1 km above sea level), using the wide-field camera called Gattini on the PLATO instrument. They found the median value of sky brightness, when the Sun elevation is less than -18$^{\circ}$ and the Moon is below the horizon, to be 22.45, 21.40, and 20.56 mag arcsec$^{-2}$ in $B$, $V$ and $R$ respectively. 

This paper presents optical night sky background levels measured from the stratosphere from the Super-pressure Balloon-borne Imaging Telescope (\superbit{}). \superbit{} is a diffraction-limited, wide-field, 0.5 m telescope capable of taking science observations with 50 milliarcsecond pointing stability from stratospheric altitudes on a balloon-borne platform. The paper is organized as follows. In \textsection{}\ref{sec:data}, we present details of the three \superbit{} commissioning flights from which we used the data for the sky background measurements. In \textsection{}\ref{sec:analysis}, we present the data analysis procedure, specifically the photometric calibration (\textsection{}\ref{ssec:photometry}), the sky brightness estimation procedure in units of ADU/s (\textsection{}\ref{ssec:bkg_adu}), and the sky brightness estimation procedure in physical units (\textsection{}\ref{ssec:bkg_physical}). In \textsection{}\ref{sec:results}, we present the results. 

\section{Data} \label{sec:data}
 The sky backgrounds were estimated using CCD images in different bands from three different commissioning flights of \superbit{} in 2016, 2018, and 2019. The 2016 flight was launched from the \textit{Columbia Scientific Balloon Facility} (CSBF-NASA) located in Palestine, Texas for a single night on June 30, 2016, and the average altitude at science observations of $\sim 34$ km. The 2016 telescope was an engineering telescope with a modified-Dall-Kirkham $f$/10 design with a 500 mm aperture. The CCD consisted of 6576 (H) $\times$ 4384 (V) pixels with a 5.5 $\mu$m $\times$ 5.5 $\mu$m pixel size, and a 0.226$^{\prime\prime}$/pixel plate scale. The 2018 flight launched on June 6, 2018 for a single night  from CSBF-NASA in Palestine, Texas \citep{Romualdez18}. The average altitude during science observations was $\sim$ 29 km. The 2018 telescope and the CCD were the same as the 2016 flight. The 2019 flight launched on September 18, 2019 for a single night from the \textit{Timmins Stratospheric Balloon Base} in Ontario, Canada, with launch support provided by the \textit{Centre National d'\'Etudes Spatiales} (CNES) and the \textit{Canadian Space Agency} (CSA). The average altitude at which science observations were taken was $\sim 34$ km. Compared to the 2016 and 2018 flights, both the telescope and the CCD were upgraded in 2019. The 2019 telescope was a science-quality telescope with a modified-Dall-Kirkham $f$/11 design also with a 500 mm aperture. The CCD was upgraded to one with improved quantum efficiency, 6576 (Horizonal) $\times$ 4384 (Vertical) pixels with a 5.5 $\mu$m $\times$ 5.5 $\mu$m pixel size, and a 0.206$^{\prime\prime}$/pixel plate scale. We refer the reader to \citet{bit2019} for further details on the \superbit{} 2019 commissioning flight.

\section{Sky background analysis} \label{sec:analysis}
\subsection{Photometry}\label{ssec:photometry}
\begin{figure*}[htb!]
    \centering
    \includegraphics[width=35em]{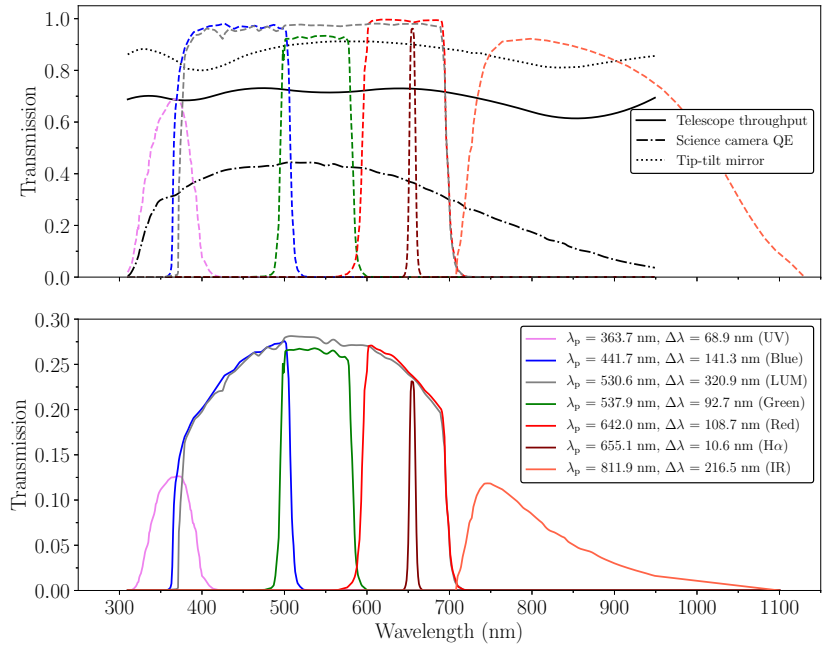}
    \caption{The overall \superbit{} bandpass (lower panel) was constructed by taking the dot product of the telescope throughput, the reflectance of the tip-tilt mirror (which is coated with protected aluminum), the transmission of the filters (dashed lines), the quantum efficiency of the science camera CCD sensor (upper panel). This figure shows the bandpass, pivot wavelengths, and the bandwidths for the 2019 flight.}
    \label{fig:sb_bp}
\end{figure*}

To estimate the night sky brightness level in physical units, it is necessary to consider the bandpass of the instrument. The \superbit{} bandpass is derived from the combination of the throughput of the telescope, the quantum efficiency of the CCD sensor, the reflectance of the tip-tilt mirror (which is coated with protected aluminum), and the transmission of the filters. The bandpass for the 2019 flight is shown in Figure \ref{fig:sb_bp}. 

To estimate the band centre, we used the source-independent pivot wavelength defined as \citep{Koornneef}

\begin{equation}
    \lambda_{\rm p} = \sqrt{\frac{\int R(\lambda)\, \lambda \, d\lambda}{\int R(\lambda) \, d\lambda / \lambda}}
\end{equation}

\noindent where $R(\lambda)$ is the bandpass response function. The bandwidth was estimated using the Kraus formula

\begin{equation}
    \Delta \lambda = \frac{\big[\int f_{\lambda}(\lambda) \, R(\lambda) \, d\lambda \big]^{2}}{\int [f_{\lambda}(\lambda)  \, R(\lambda)]^{2} \, d\lambda}
\end{equation}

\noindent where $f_{\lambda}(\lambda)$ is the flux density of the source for which we assumed a flat-spectrum. The pivot wavelengths and bandwidths for the \superbit{} 2016, 2018, and 2019 flights are given in Table \ref{tab:bandpass_wavelengths}. The values for the standard Johnson-Cousins \textit{UBVRI} system are also shown for comparison \citep{Bessell_2012}. The bandwidth for the  \textit{UBVRI} system in Table \ref{tab:bandpass_wavelengths} is the full width at half maximum (FWHM).

\begin{deluxetable}{|c|r|c|c|c|c|c|c|}
\tabletypesize{\small}
\tablecolumns{10}
\setlength\tabcolsep{1.8pt}
\label{tab:bandpass_wavelengths}
\tablecaption{The pivot wavelengths and bandwidths for the \superbit{}  2016, 2018, and 2019 flights as well as the standard Johnson-Cousins \textit{UBVRI} system  \citep{Bessell_2012} are shown for comparison.}
\tablehead{
\colhead{Year} & \colhead{Filter} & \colhead{Lum} & \colhead{UV} & \colhead{Blue} & \colhead{Green} & \colhead{Red} & \colhead{IR}
}
\startdata
2016, 2018 & $\lambda_{\rm p}$ (nm) & 519.3 & 365.5 & 442.1 & 536.6 & 640.0 & 809.7 \\
2016, 2018 & $\Delta\lambda$ (nm) & 312.2 & 67.6 & 140.7 & 92.2 & 107.7 & 211.9 \\ \hline
2019 & $\lambda_{\rm p}$ (nm) & 530.6 & 363.7 & 441.7 & 537.9 & 642.0 & 811.9 \\
2019 & $\Delta\lambda$ (nm) & 320.9 & 68.9 & 141.3 & 92.7 & 108.7 & 216.5 \\ \hline
\nodata & \multicolumn{1}{c|}{Filter} & \nodata & $U$ & $B$ & $V$ & $R$ & $I$ \\ \hline
\nodata & $\lambda_{\rm p}$ (nm) & \nodata & 359.7 & 437.7 & 548.8 & 651.5 & 798.1 \\
\nodata & $\Delta\lambda$ (nm) & \nodata & 62.5 & 89.0 & 83.0 & 144.3 & 149.9 \\ 
\enddata
\end{deluxetable}

\begin{figure*}[htb!]
    \centering
    \includegraphics[width=30em]{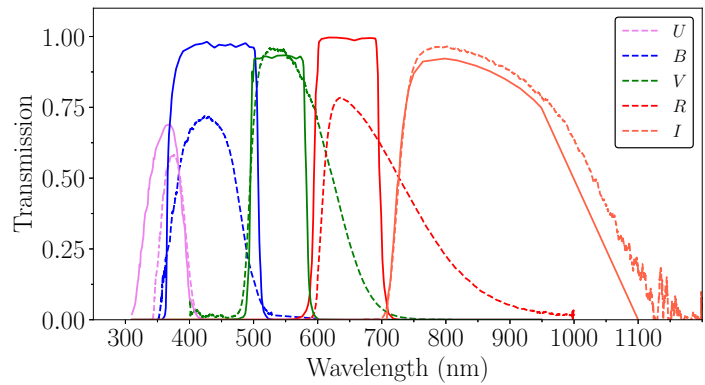}
    \caption{Comparison of the \superbit{} filters and the Johnson-Cousins filters on the Gemini North Acquisition Camera. The solid lines are the \superbit{} filters and dashed lines are the Gemini North filters. There is reasonable overlap between $U$ and UV, $B$ and Blue, $V$ and Green, $R$ and Red, and $I$ and IR between the Johnson-Cousins filters and the \superbit{} filters, respectively. }
    \label{fig:filter_compare}
\end{figure*}

 Figure \ref{fig:filter_compare} shows the overlap of the \superbit{} filters and the Johnson-Cousins filters on the Gemini North Acquisition Camera taken from the Spanish Virtual Observatory Filter Profile Service\footnote{http://svo2.cab.inta-csic.es/theory/fps/}. There is reasonable overlap between $U$ and UV, $B$ and Blue, $V$ and Green, Red and $R$, and $I$ and IR between the Johnson-Cousins and \superbit{} filters, respectively.

\subsection{Sky background estimation in ADU/s}\label{ssec:bkg_adu}
The raw CCD images were bias, dark current, and cosmic-ray corrected. To estimate the background level in units of ADU/s, pixel values $\pm3\sigma$ away from the mean of reduced image were discarded until convergence, where the final iteration clips no pixels. The remaining ($\pm3\sigma$ clipped) pixels were fit with a Gaussian distribution. The estimate of the sky background level was taken to be the mean of the Gaussian fit. The error in the sky background level in ADU/s was taken to be the error in the mean, which was calculated using the bootstrap method. For a given $N$ number of pixels that remain after $\pm3\sigma$ clipping, the bootstrap method for estimating the error in the mean consisted of the following steps: 

\begin{enumerate}
    \item Take a random sample of $N$ pixels with replacement
    \item Take the mean of the random sample
    \item Repeat steps (1) and (2) for $M$ = 5000 iterations
    \item Take the standard deviation of the sample of $M$ means to estimate the error in the mean
\end{enumerate}

\begin{figure*}[htb!]
    \centering
    \includegraphics[width=45em]{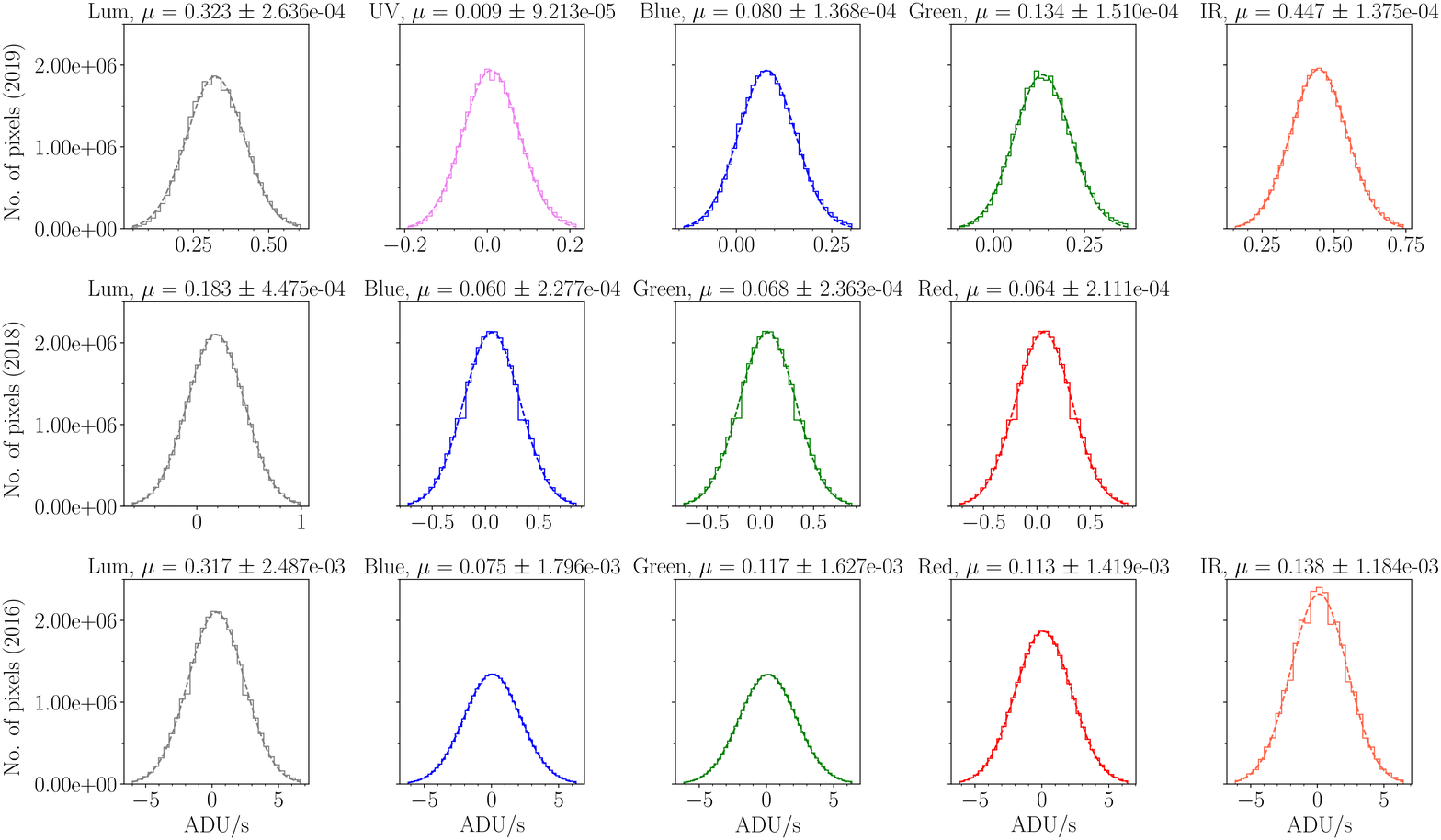}
    \caption{The mean sky background level and the error in the mean in ADU/s. The error in the mean in the background level was estimated using the bootstrap method. The results are shown for three commissioning flights of \superbit{} in 2016, 2018, and 2019. The exposure times for the images used were 20 s, 120 s, and 300 s for 2016, 2018, and 2019, respectively.} 
    \label{fig:adu_per_sec}
\end{figure*}

To test whether additional masking of any residual diffuse emission from galaxies after the $\pm 3\sigma$ clipping would be necessary, we compared the mean and the error in the mean in the background level using the 2018 Lum image between two cases: (i) the $\pm 3\sigma$ clipped image; (ii) the $\pm 3\sigma$ clipped image with additional masking of residual diffuse emission from galaxies. We found that the means of the Gaussian distributions between the two cases were identical, and the difference between the errors in the means was $< 2\%$. Therefore, we concluded that $\pm3\sigma$ clipping of the reduced image is sufficient for the purpose of sky background estimation. The mean sky background level and its error in ADU/s for the three different years and different bands are shown in Figure \ref{fig:adu_per_sec}. The exposure times for the images taken were 20 s, 120 s, and 300 s for 2016, 2018, and 2019, respectively. 

\subsection{Sky background estimation in physical units} \label{ssec:bkg_physical}
\begin{figure*}[htb!]
    \centering
    \includegraphics[width=15cm]{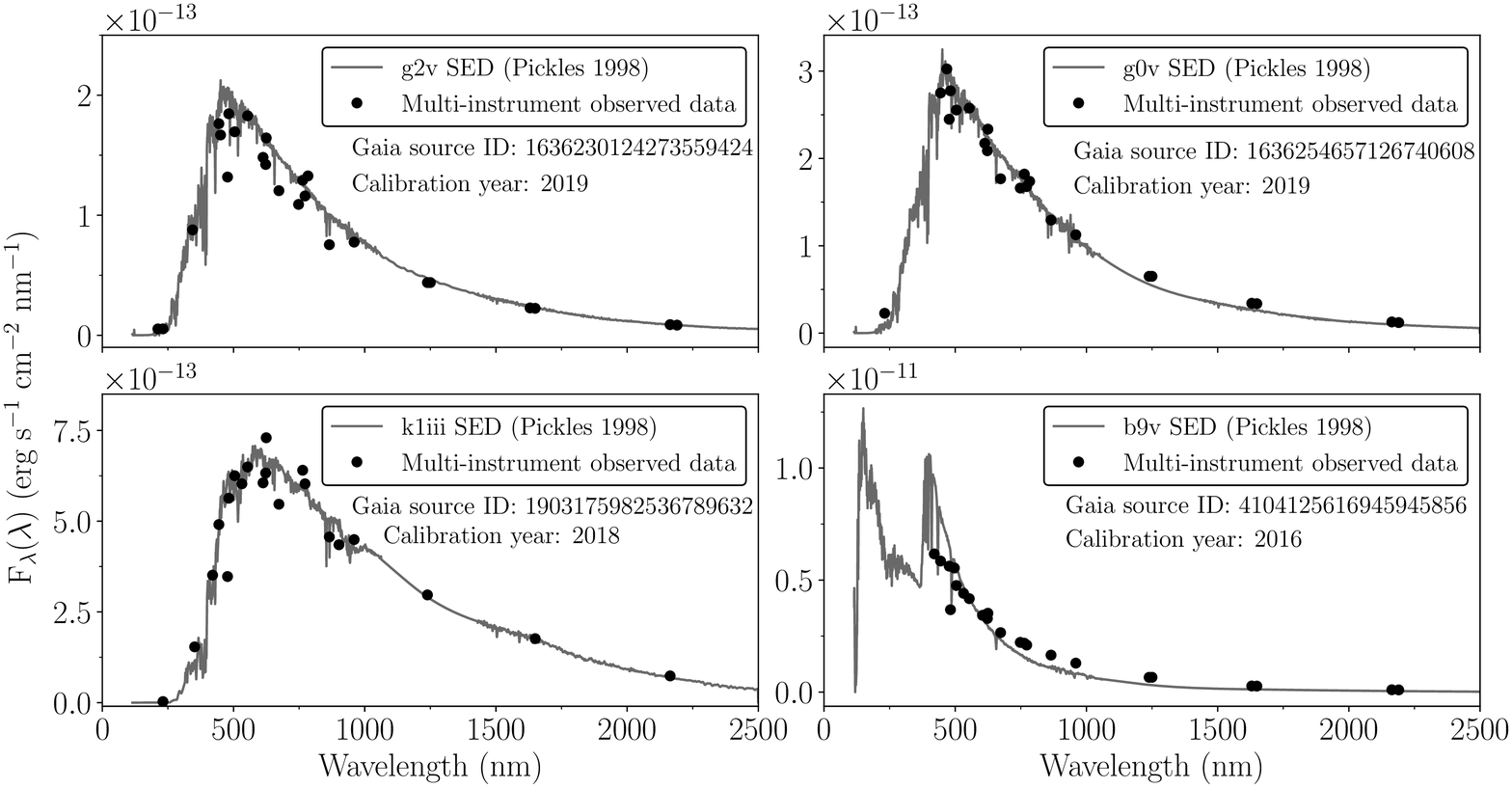}
    \caption{Spectral type estimates for the calibration stars with template fitting for 2016, 2018, and 2019 data. The spectral templates were taken from the stellar spectral flux library by \citet{pickles}.}
    \label{fig:templates}
\end{figure*}

To convert the background estimate from ADU to physical units, the dot product between the spectral energy distribution (SED) of an unsaturated calibrator star and the bandpass was first taken. The spectral type of the calibrator stars was estimated by extracting observed optical flux as a function of wavelength data points measured by other instruments for the calibrator star within a circle of radius 2 arcseconds using the \texttt{VizieR photometry tool}\footnote{http://vizier.unistra.fr/vizier/sed/}. The observed data points were taken from Pan-STARRS DR1 \citep{ps1}, Gaia DR2 \citep{gaia_dr2}, AAVSO Photometric All-Sky Survey (APASS) \citep{aavso}, 2MASS All-Sky Catalog \citep{2mass}, Guide Star Catalog 2.3.2 \citep{GSC}, and the UCAC5 catalogue \citep{ucac5}. The measured data were least-squares fit to stellar SED templates from a standard stellar spectra flux library by \citet{pickles} to estimate the spectral type of the calibrator star. The SED and the best-fit to observed data for the calibration stars used for the 2016, 2018, and 2019 data are shown in Figure \ref{fig:templates}. 

The Gaia DR2 catalogue was used for external flux calibration of the \superbit{} data. Gaia DR2 magnitudes are defined by 

\begin{equation}
    G = -2.5\log_{10}\bar{I} + G_{0}
\end{equation}

\noindent where $\bar{I}$ is the internally calibrated flux in units of photo-electrons/s, and $G_{0}$ is the zero-point, which is provided by Gaia DR2 in both the Vega and AB magnitude systems. Throughout this paper, we use the AB magnitude system, defined such that a source with a flux density $f_{\nu}$ of $3.631 \times 10^{-20}$ erg s$^{-1}$ cm$^{-2}$ Hz$^{-1}$ has $m_{\rm AB} = 0$.

\begin{equation}
    m_{\rm AB} = -2.5 \log_{10}f_{\nu} - 48.60
\end{equation}

\noindent The theoretical flux of the calibrator star is calculated on the Gaia scale. The proper normalization of the SED, $S(\lambda)$, was then determined given the observed Gaia BP band flux. $S(\lambda)$ data from \citet{pickles} is in units of erg s$^{-1}$ cm$^{-2}$ \AA$^{-1}$ and is arbitrarily normalized at $\lambda = 555.6$ nm. $S(\lambda)$ was renormalized by comparing the theoretical and observed flux. To do so, the theoretical integrated flux was first converted to units of photo-electrons/s \citep{gaia_photo_1, gaia_photo_2}. 

\begin{equation}
    \bar{I} \equiv  I_{\rm obs, Gaia, *} = \beta \, I_{\rm th, Gaia, *}
\end{equation}

\begin{equation}
\begin{split}
 \beta \,I_{\rm th, Gaia, *} =  \frac{P_{\rm A}} {hc} \int_{\lambda = 0}^{\infty} \beta S(\lambda) \cdot R_{\rm BP}(\lambda) \cdot \lambda \,d\lambda
\end{split} \label{gaia_sed}
\end{equation}

\noindent where $P_{\rm A} = 0.7278$ m$^{2}$ is the Gaia telescope pupil area, $R_{\rm BP}(\lambda)$ is the Gaia BP bandpass, and $\beta$ is the renormalization factor for $S(\lambda)$. Once $\beta$ is calculated, the observed flux density of the calibrator star on the \superbit{} flux scale is then

\begin{equation}
    f_{\nu, \rm SB, *} = \frac{\int \beta S(\lambda) \cdot R_{\rm SB}(\lambda)\, d\lambda}{\int R_{\rm SB}(\lambda) \cdot \frac{c}{\lambda^{2}}\, d\lambda }\rm \;\;\; \Bigg[\frac{erg}{s}\frac{1}{cm^{2}}\frac{1}{Hz}\Bigg]
\end{equation}

\noindent where the numerator is the observed integrated flux and after normalization for the bandpass, $f_{\nu, \rm SB, *}$ is the observed flux density of the calibrator star. 

With the observed flux density, an ADU/s to flux density conversion factor was calculated. This scale factor provides an indication of the sensitivity of the instrument. The ADU/s for the calibrator star were taken using the automatic aperture photometry routine of \texttt{SExtractor}, which is derived from Kron's first moment algorithm \citep{Kron, sextractor}. This was done after the reduced (bias, dark, and cosmic-ray corrected) image was also background subtracted. Once calculated, the flux density conversion factor, $\alpha$, is assumed to be valid over the entire image and was used to convert the sky background level from ADU/s to physical units.

\begin{equation}
    \alpha_{\rm SB, *} \equiv \alpha = \frac{\rm (ADU/s)_{SB, *}}{f_{\nu, \rm SB, *}}
\end{equation}

We first converted the sky background level in ADU/s/pixel to ADU/s/arcsec$^{2}$ given the CCD pixel scale. The background level in ADU/s/arcsec$^{2}$ was then converted to a flux density per arcsec$^{2}$ and subsequently to $m_{\rm AB, bkg / arcsec ^{2}}$ by 

\begin{equation}
      f_{\rm \nu, bkg / arcsec^{2}} = \frac{\rm ADU/s} {\alpha} 
\end{equation}

\begin{equation}
    m_{\rm AB, bkg / arcsec^{2}}  = -2.5 \log_{10}f_{\rm \nu, bkg / arcsec^{2}} - 48.60 
    \label{AB_mag_eqn}
\end{equation}

To estimate the error in the background level in physical units, we ran 2000 Monte Carlo (MC) simulations, for which random samples were drawn assuming a Gaussian distribution for parameters that go into the calculation of the sky background. Table \ref{tab:mc_parameters} lists the parameters that were sampled in the MC simulations and how the errors in the parameters were obtained. The sky background level and its error in physical units were then taken to be the mean and the standard deviation of the Gaussian fit to the results from the MC simulations (see Figure \ref{fig:mc_bkg_asec}). 

\begin{deluxetable*}{|c|c|l|}
\tabletypesize{\small}
\tablecolumns{10}
\setlength\tabcolsep{1.4pt}
\label{tab:mc_parameters}
\tablecaption{List of parameters sampled in the Monte Carlo simulations. An example of the relative uncertainty of the sampled parameters for the 2019 Lum calibration are shown in Column 2.}
\tablehead{
\multicolumn{1}{c}{Parameter} & \multicolumn{1}{c}{Example (2019 Lum calibration)} & \multicolumn{1}{c}{Description}
}
\startdata
$I_{\rm obs, Gaia, *}$ & 56170.18 $\pm$ 45.91 [e$^{-}$/s] & Observed BP flux for the calibrator star provided by Gaia DR2.\\
$R_{\rm BP} (\lambda)\tablenotemark{\scriptsize a}$ & 0.65 $\pm$ 7.99E-4 [dimensionless] & BP bandpass provided by Gaia DR2.\\
$S(\lambda)$\tablenotemark{\scriptsize b} & 1.055 $\pm$ 0.007 [erg s$^{-1}$ cm$^{-2}$ \AA$^{-1}$] & Calibrator star SED provided by \citet{pickles} stellar spectral flux library.\\
(ADU/s)$_{\rm SB, *}$ & 13574.09 $\pm$ 2.78 [ADU/s] & Taken from the automatic aperture photometry routine by \texttt{SExtractor}.\\
(ADU/s)$_{\rm bkg/arcsec^{2}}$ & 0.32 $\pm$ 2.64E-4 [ADU/s] & Taken as the the mean of the Gaussian distribution in Figure \ref{fig:adu_per_sec}. 
\enddata
\tablenotetext{a}{The bandpass value provided in Column 2 is at the Gaia BP pivot wavelength of 505.15 nm.}
\tablenotetext{b}{The SED value provided in Column 2 is at the pivot wavelength (530.6 nm) of the \superbit{} 2019 Lum band. Note that the SED data from \citet{pickles} is arbitrarily normalized at $\lambda = 555.6$ nm.}
\end{deluxetable*}

\section{Results}\label{sec:results}

\begin{deluxetable*}{|c|c|c|c|c|c|c|c|c|c|}
\tabletypesize{\scriptsize}
\tablecolumns{10}
\setlength\tabcolsep{1.4pt}
\label{tab:calibration_all}
\tablecaption{Photometric calibration parameters (for different bands and years). The exposure times for the images were 20 s, 120 s, and 300 s for 2016, 2018, and 2019, respectively.}
\tablehead{
\colhead{Obs. time} & \colhead{Band} & \colhead{$\lambda_{\rm p}$\tablenotemark{\scriptsize a}} & \colhead{ADU$_{\rm bkg}$\tablenotemark{\scriptsize b}} & \colhead{$\alpha$\tablenotemark{\scriptsize c}} & \colhead{$\beta$\tablenotemark{\scriptsize d}} & \colhead{Gaia source ID\tablenotemark{\scriptsize e}} & \colhead{Sp.\tablenotemark{\scriptsize f}} & \colhead{Gaia\tablenotemark{\scriptsize g}} & \colhead{\superbit{}\tablenotemark{\scriptsize h}}  \\[-0.2cm]
\colhead{(Local)} & \colhead{} & \colhead{(nm)} & \colhead{per s} & \colhead{(ADU s $^{-1}$ erg$^{-1}$ cm$^{2}$)} & \colhead{(dimensionless)} & \colhead{} & \colhead{type} & \colhead{BP mag} & \colhead{mag}
}
\startdata
2019-09-18, 04:02:48     & Lum     & 530.6     & 0.32 $\pm$ 2.64E-4  & 9.07E28 $\pm$ 5.34E25 & 1.79E-14 $\pm$ 3.24E-17   & 1636230124273559424    & g2v      & 13.507 $\pm$ 0.001 & 13.462 $\pm$ 0.001     \\
2019-09-18, 05:14:39     & UV      & 363.7     & 0.01 $\pm$ 9.21E-5  & 1.76E27 $\pm$ 1.71E25 & 2.65E-14 $\pm$ 5.14E-17   &  1636254657126740608   & g0v      & 13.059 $\pm$ 0.001 & 14.243 $\pm$ 0.004      \\
2019-09-18, 05:07:59     & Blue    & 441.7     & 0.08 $\pm$ 1.37E-4  & 4.22E28 $\pm$ 8.39E25 & 2.65E-14 $\pm$ 5.14E-17   & 1636254657126740608    & g0v      & 13.059 $\pm$ 0.001 & 13.313 $\pm$ 0.002      \\
2019-09-18, 05:29:42      & Green   & 537.9     & 0.13 $\pm$ 1.51E-4  & 2.89E28 $\pm$ 2.22E25 & 2.65E-14 $\pm$ 5.14E-17   & 1636254657126740608    & g0v      & 13.059 $\pm$ 0.001 & 12.864 $\pm$ 0.001      \\ 
\nodata      & Red   & \nodata     & \nodata  & \nodata & \nodata  & \nodata    & \nodata      & \nodata & \nodata     \\ 
2019-09-18, 05:22:46     & IR      & 811.9     & 0.45 $\pm$ 1.38E-4  & 1.05E28 $\pm$ 1.03E25 & 2.65E-14 $\pm$ 5.14E-17   & 1636254657126740608    & g0v      & 13.059 $\pm$ 0.001 & 12.563 $\pm$ 0.001      \\ 
\hline
2018-06-06, 02:59:37     & Lum     & 519.3     & 0.18 $\pm$ 4.48E-4  & 5.48E28 $\pm$ 3.64E25 & 7.20E-14 $\pm$ 1.65E-16   & 1903175982536789632    & k1iii    & 12.087 $\pm$ 0.001 & 12.097 $\pm$ 0.001      \\
\nodata      & UV   & \nodata     & \nodata  & \nodata & \nodata  & \nodata    & \nodata      & \nodata & \nodata     \\ 
2018-06-06, 03:28:52     & Blue    & 442.1     & 0.06 $\pm$ 2.28E-4  & 4.03E28 $\pm$ 5.89E25 & 7.20E-14 $\pm$ 1.65E-16   & 1903175982536789632    & k1iii    & 12.087 $\pm$ 0.001 & 12.677 $\pm$ 0.002      \\
2018-06-06, 03:20:28      & Green   & 536.6     & 0.07 $\pm$ 2.36E-4  & 2.08E28 $\pm$ 2.04E25 & 7.20E-14 $\pm$ 1.65E-16   & 1903175982536789632    & k1iii    & 12.087 $\pm$ 0.001 & 11.843 $\pm$ 0.001      \\
2018-06-06, 03:06:23     & Red     & 640.0     & 0.06 $\pm$ 2.11E-4  & 1.24E28 $\pm$ 8.98E24 & 7.20E-14 $\pm$ 1.65E-16   & 1903175982536789632    & k1iii    & 12.087 $\pm$ 0.001 & 11.406 $\pm$ 0.001      \\
\nodata      & IR   & \nodata     & \nodata  & \nodata & \nodata  & \nodata    & \nodata      & \nodata & \nodata     \\ 
 \hline
 2016-07-01, 01:04:10      & Lum     & 519.3       & 0.32 $\pm$ 2.49E-3  & 5.32E28 $\pm$ 8.18E25 & 3.93E-13 $\pm$ 1.26E-15   & 4104125616945945856    & b9v      & 9.884  $\pm$ 0.003  & 9.811 $\pm$ 0.002       \\
 \nodata      & UV   & \nodata     & \nodata  & \nodata & \nodata  & \nodata    & \nodata      & \nodata & \nodata     \\ 
 2016-07-01, 12:51:00      & Blue    & 442.1       & 0.08 $\pm$ 1.80E-3  & 3.41E28 $\pm$ 8.12E25 & 3.93E-13 $\pm$ 1.26E-15   & 4104125616945945856    & b9v      & 9.884  $\pm$ 0.003  & 9.763 $\pm$ 0.003       \\
 2016-07-01, 12:49:41     & Green   & 536.6      & 0.12 $\pm$ 1.63E-3  & 2.44E28 $\pm$ 4.07E25 & 3.93E-13 $\pm$ 1.26E-15   & 4104125616945945856    & b9v      & 9.884  $\pm$ 0.003  & 9.839 $\pm$ 0.002       \\
 2016-07-01, 12:47:30      & Red     & 640.0      & 0.11 $\pm$ 1.42E-3  & 1.58E28 $\pm$ 2.97E25 & 3.93E-13 $\pm$ 1.26E-15   & 4104125616945945856    & b9v      & 9.884  $\pm$ 0.003 & 10.027 $\pm$ 0.002      \\
2016-07-01, 12:44:20      & IR      & 809.7      & 0.14 $\pm$ 1.18E-3 & 7.29E27 $\pm$ 1.77E25 & 3.93E-13 $\pm$ 1.26E-15   & 4104125616945945856    & b9v      & 9.884  $\pm$ 0.003 & 10.290 $\pm$ 0.001  
\enddata
\tablenotetext{a}{Pivot wavelength of the band (nm).}
\tablenotetext{b}{Sky background level in raw units of ADU/s.}
\tablenotetext{c}{Raw count rate (ADU/s) to flux density (erg s$^{-1}$ cm$^{-2}$ Hz$^{-1}$) conversion factor. This factor provides an estimate for the sensitivity of the instrument per band.}
\tablenotetext{d}{Dimensionless renormalization factor for stellar spectral energy distribution template from \citep{pickles}.}
\tablenotetext{e}{Gaia DR2 source ID of calibrator star.}
\tablenotetext{f}{Spectral type of calibrator star.}
\tablenotetext{g}{Gaia DR2 magnitude in the BP band of the calibrator star.}
\tablenotetext{h}{\superbit{} magnitude of the calibrator star. }
\end{deluxetable*}

\begin{longrotatetable}
\begin{deluxetable*}{|c|c|c|c|c|c|c|c|c|c|c|c|c|c|c|c|c|c|c|c|}
\tabletypesize{\scriptsize}
\tablecolumns{20}
\label{bkg_all_years}
\setlength\tabcolsep{1pt}
\tablecaption{Sky background estimates from the stratosphere from balloon-borne observations with the \superbit{} telescope.}
\tablehead{
    \colhead{Obs. time} & \colhead{Type} & \colhead{Band} & \colhead{$\lambda_{\rm p}$\tablenotemark{\scriptsize a}} & \colhead{$m_{\rm AB, bkg}$\tablenotemark{\scriptsize b}} & \colhead{$m_{\rm AB, bkg}$\tablenotemark{\scriptsize c}} & \colhead{$f_{\nu, \rm bkg}$\tablenotemark{\scriptsize d}} & \colhead{$f_{\nu, \rm ZL}$\tablenotemark{\scriptsize e}} & \colhead{$f_{\nu, \rm DGL+EBL}$\tablenotemark{\scriptsize f}} & \colhead{Alt.\tablenotemark{\scriptsize g}} & \colhead{$a$\tablenotemark{\scriptsize h}} & \colhead{Moon\tablenotemark{\scriptsize i}} & \colhead{$\ell$\tablenotemark{\scriptsize j}} & \colhead{$b$\tablenotemark{\scriptsize k}} & \colhead{$\lambda - \lambda_{\odot}$\tablenotemark{\scriptsize l}} & \colhead{$\beta$\tablenotemark{\scriptsize m}} & \colhead{$\lambda_{0}$\tablenotemark{\scriptsize n}} & \colhead{$\phi_{0}$\tablenotemark{\scriptsize o}} & \colhead{$N_{\rm m}$\tablenotemark{\scriptsize p}} & \colhead{$a_{\odot}$\tablenotemark{\scriptsize q}} \\ 
\colhead{(Local)} & \colhead{} & \colhead{} & \colhead{(nm)} & \colhead{(per arcsec$^{2}$)} & \colhead{(per pixel)}  & \colhead{($\mu$Jy/arcsec$^{2}$)} & \colhead{($\mu$Jy/arcsec$^{2}$)} & \colhead{($\mu$Jy/arcsec$^{2}$)} & \colhead{(km)} & \colhead{($^{\circ}$)} & \colhead{($^{\circ}$)} & \colhead{($^{\circ}$)} & \colhead{($^{\circ}$)}  & \colhead{($^{\circ}$)}  &  \colhead{($^{\circ}$)} & \colhead{($^{\circ}$)} & \colhead{($^{\circ}$)} & \colhead{} & \colhead{($^{\circ}$)} 
}
\startdata
2019-09-18, 04:02:48 & $I_{\rm bkg, raw}$\tablenotemark{\scriptsize r} & Lum   & 530.6 &  21.594 $\pm$ 0.001 & 25.025 $\pm$ 0.001 & 8.364 $\pm$ 0.008   & 1.679  $\pm$ 0.168 & 0.273 $\pm$ 0.068 & 34.25 & 26.16 & 100.63 & 97.72 & 38.10   & 61.75 & 81.64 & -81.90   & 47.14 & 10 & -29.62\\
 & $I_{\rm ZL}$\tablenotemark{\scriptsize s} &  &  &  21.837 $\pm$ 0.027 & 25.268 $\pm$ 0.027 & 6.684 $\pm$ 0.168  & &  &  &  &  & & &  &   &  &  &  & \\
& $I_{\rm DGL}$\tablenotemark{\scriptsize t} &  &  &  21.883 $\pm$ 0.031 & 25.313 $\pm$ 0.031 & 6.411 $\pm$ 0.181  & & &  &  &  &  & & &  &  &  &   & \\
& $I_{\rm A}$ to zenith\tablenotemark{\scriptsize u} &  &  &  22.748 $\pm$ 0.031 & 26.179 $\pm$ 0.031 & 2.888 $\pm$ 0.082  & &  &  &  &  & & &  &   &  & &  & \\ 
2018-06-06, 02:59:37 & $I_{\rm bkg, raw}$ & Lum & 519.3 & 21.852 $\pm$ 0.001 & 25.081 $\pm$ 0.001 & 6.595 $\pm$ 0.006 & 2.089 $\pm$ 0.209   &  0.546 $\pm$ 0.109  & 28.93 & 38.60  & 45.47  & 93.79 & -20.72 & 96.72 & 39.41  & -97.30  & 31.61 & 7  & -31.45\\
& $I_{\rm ZL}$ &  &  &  22.266 $\pm$ 0.050 & 25.495 $\pm$ 0.050 & 4.506 $\pm$ 0.209  &  &  &  & &  & & &  &   &  &  &  &\\
& $I_{\rm DGL}$   &  &  &  22.406 $\pm$ 0.065 & 25.635 $\pm$ 0.065 & 3.959 $\pm$ 0.236  &  &  &  & &  & & & &  &   &  &  &  \\
& $I_{\rm A}$ to zenith &  &  &  22.911 $\pm$ 0.065 & 26.140 $\pm$ 0.065 & 2.488 $\pm$ 0.148  &  &  & &  &  & & &  &   & & &  &  \\ 
 2016-07-01, 01:04:13 & $I_{\rm bkg, raw}$ & Lum  & 519.3 & 21.201 $\pm$ 0.002 & 24.430 $\pm$ 0.002 & 12.012 $\pm$ 0.022 & 3.468 $\pm$ 0.347 &  0.956 $\pm$ 0.191 & 34.14 & 43.54 & 139.55 & 17.56 & -1.92  & 144.32  & 8.71  & -100.37 & 31.40  & 3  & -34.65\\
& $I_{\rm ZL}$ &  &  &  21.571 $\pm$ 0.044 & 24.800 $\pm$ 0.044 &  8.543 $\pm$ 0.348 &  &  &  &  &  & & &  &   &  &  &  &\\
& $I_{\rm DGL}$  &  &  &  21.700 $\pm$ 0.057 & 24.929 $\pm$ 0.057 &  7.588 $\pm$ 0.397 &  &  &  &  &  & & & &  &   &  &  &  \\
& $I_{\rm A}$ to zenith &  &  &  22.098 $\pm$ 0.057 & 25.328 $\pm$ 0.057 & 5.257 $\pm$ 0.275 &  &  & &  &  & & &  &   &  & & & \\ \hline 
2019-09-18, 05:14:39 & $I_{\rm bkg, raw}$ & UV  & 363.7  & 21.177 $\pm$ 0.011 & 24.608 $\pm$ 0.011 & 12.280 $\pm$ 0.124  & 0.419 $\pm$ 0.042 & 0.068 $\pm$ 0.017 & 33.42 & 23.75 & 100.54 & 97.68 & 38.12  & 61.96 & 81.64 & -82.30   & 46.97 & 10 & -19.81 \\
& $I_{\rm ZL}$ &  &  &  21.215 $\pm$ 0.012 & 24.645 $\pm$ 0.012 & 11.861 $\pm$ 0.131 & & &  &  &  & & &  &   &  &  &  &\\ 
& $I_{\rm DGL}$  &  &  &  21.221 $\pm$ 0.012 & 24.652 $\pm$ 0.012 & 11.793 $\pm$ 0.132 & & & &  &  &  & & & &     &  &  &  \\ 
& $I_{\rm A}$ to zenith &  &  &  22.174 $\pm$ 0.012 & 25.604 $\pm$ 0.012 & 4.903 $\pm$ 0.055  & & & &  &  &  & & &  &   & &  & \\ \hline 
2019-09-18, 05:07:59 & $I_{\rm bkg, raw}$ & Blue  & 441.7 &  22.279 $\pm$ 0.002 & 25.710 $\pm$ 0.002 & 4.450 $\pm$ 0.008 & 1.111 $\pm$ 0.111 & 0.181 $\pm$ 0.045 & 33.54 & 23.90  & 100.55 & 97.68 & 38.12  & 61.98  & 81.64 & -82.26  & 46.98 & 10 & -20.80\\
 & $I_{\rm ZL}$ &  &  &  22.591 $\pm$ 0.036 & 26.021 $\pm$ 0.036 & 3.340 $\pm$ 0.111 &  &  &  &  &  & & &  &   &  &  &  & \\
 & $I_{\rm DGL}$   &  &  &  22.651 $\pm$ 0.041 & 26.082 $\pm$ 0.041 & 3.159 $\pm$ 0.120 & & &  &  &  &  & & &  &   &  &  &  \\
& $I_{\rm A}$ to zenith &  &  &  23.604 $\pm$ 0.041 & 27.035 $\pm$ 0.041 & 1.313 $\pm$ 0.050  & &  &  &  &  & & &  &  &  &  & &  \\  
2018-06-06, 03:28:52 & $I_{\rm bkg, raw}$ & Blue & 442.1 & 22.725 $\pm$ 0.002 & 25.954 $\pm$ 0.002 & 2.951 $\pm$ 0.005 & 1.382 $\pm$ 0.138 & 0.361 $\pm$ 0.072 & 28.44 & 44.38 & 45.43  & 93.79 & -20.72 & 96.71 & 39.41  & -97.40   & 31.61 & 7 & -28.53  \\
& $I_{\rm ZL}$ &  &  &  23.411 $\pm$ 0.096 & 26.640 $\pm$ 0.096 & 1.569 $\pm$ 0.138  &  &  &  &  & & &  &   &  & & & & \\
& $I_{\rm DGL}$  &  &  &  23.695 $\pm$ 0.140 & 26.924 $\pm$ 0.140 & 1.208 $\pm$ 0.156  &  &  &  &  & &  & &  &   &  & & &  \\
& $I_{\rm A}$ to zenith &  &  &  24.078 $\pm$ 0.140 & 27.307 $\pm$ 0.140 & 0.849 $\pm$ 0.110  &  &  &  &  & & &  &   &  & & & & \\ 
 2016-07-01, 12:51:05 & $I_{\rm bkg, raw}$ & Blue & 442.1  & 22.621 $\pm$ 0.003 & 25.850 $\pm$ 0.003 & 3.248 $\pm$ 0.009 & 2.294 $\pm$ 0.229 & 0.632 $\pm$ 0.126  & 34.29 & 43.01 & 139.38 & 17.56 & -1.92  & 144.32 & 8.71  & -100.14 & 31.39 & 3  & -34.06 \\
 & $I_{\rm ZL}$ &  &  &  23.951 $\pm$ 0.261 & 27.181 $\pm$ 0.261 & 0.954 $\pm$ 0.230  & &  &  &  &  & & &  &   &  &  & & \\
 & $I_{\rm DGL}$   &  &  &  25.131 $\pm$ 0.884 & 28.361 $\pm$ 0.884 & 0.322 $\pm$ 0.262  & &  &  &  &  & &  & &  &   &  &  &  \\
& $I_{\rm A}$ to zenith &  &  &  25.540 $\pm$ 0.884 & 28.770 $\pm$ 0.884 & 0.221 $\pm$ 0.180  &  &  &  &  & & &  &  & & & &  & \\ \hline 
2019-09-18, 05:29:42 & $I_{\rm bkg, raw}$ & Green & 537.9 & 21.305 $\pm$ 0.001 & 24.736 $\pm$ 0.001 & 10.914 $\pm$ 0.010  & 1.679 $\pm$ 0.168 &  0.273 $\pm$ 0.068 & 33.19 & 23.45 & 100.51 & 97.68 & 38.12  & 61.98   & 81.62 & -82.35  & 46.94 & 10  & -17.53\\ 
& $I_{\rm ZL}$ &  &  &  21.486 $\pm$ 0.020 & 24.917 $\pm$ 0.020 & 9.235 $\pm$ 0.168  &  &  &  &  & & &  &   &  & &  & & \\
& $I_{\rm DGL}$ &  &  &  21.519 $\pm$ 0.022 & 24.950 $\pm$ 0.022 & 8.962 $\pm$ 0.182  &  &  &  &  & & &  &  &   &  & &  &  \\
& $I_{\rm A}$ to zenith &  &  &  22.490 $\pm$ 0.022 & 25.921 $\pm$ 0.022 & 3.663 $\pm$ 0.074 &  &  &  &  & & &  &   &  & &  & & \\  
2018-06-06, 03:20:28 & $I_{\rm bkg, raw}$ & Green & 536.6 & 21.872 $\pm$ 0.002 & 25.101 $\pm$ 0.002 & 6.474 $\pm$ 0.012 &  2.089 $\pm$ 0.209 &  0.546 $\pm$  0.109  &  28.59 & 42.71 & 45.44  & 93.79 & -20.72 & 96.71 & 39.41  & -97.37  & 31.59 & 7 & -29.45 \\ 
& $I_{\rm ZL}$ &  &  &  22.295 $\pm$ 0.052 & 25.524 $\pm$ 0.052 & 4.385 $\pm$ 0.209  &  &  &  &  & & & &  &   &  &  &  & \\
& $I_{\rm DGL}$   &  &  &  22.439 $\pm$ 0.067 & 25.669 $\pm$ 0.067 & 3.839 $\pm$ 0.236  &  &  &  &  & & & & &  &   &  &  &  \\
& $I_{\rm A}$ to zenith &  &  &  22.855 $\pm$ 0.067 & 26.085 $\pm$ 0.067 & 2.618 $\pm$ 0.161 &  &  &  & &  & & &  &   &  & &  & \\  
 2016-07-01, 12:49:41 & $I_{\rm bkg, raw}$ & Green & 536.6 & 21.650 $\pm$ 0.002 & 24.879 $\pm$ 0.002 & 7.943 $\pm$ 0.015 &  3.468 $\pm$ 0.347 &  0.956 $\pm$ 0.191 &  34.34 & 42.95 & 139.36 & 17.56 & -1.92  & 144.32  & 8.71  & -100.12 & 31.40  & 3  & -33.99 \\
 & $I_{\rm ZL}$  &  &  &  22.273 $\pm$ 0.084 &  25.502 $\pm$ 0.084 &  4.475 $\pm$ 0.347  & &  &  &  &  & & &  &  & &  & & \\
 & $I_{\rm DGL}$  &  &  &  22.534 $\pm$ 0.122 &  25.763 $\pm$ 0.122 &  3.519 $\pm$ 0.396  & &  &  &  &  & & & &  &  & &  & \\
& $I_{\rm A}$ to zenith &  &  &  22.944 $\pm$ 0.122 & 26.173 $\pm$ 0.122  &  2.413 $\pm$ 0.272  &  &  &  &  & & &  & &   & &  &  & \\ \hline 
2018-06-06, 03:06:23 & $I_{\rm bkg, raw}$ & Red & 640.0 & 21.385 $\pm$ 0.001 & 24.614 $\pm$ 0.001 & 10.139 $\pm$ 0.009 & 2.580 $\pm$ 0.258 &  0.674 $\pm$ 0.135 &  28.72 & 39.92 & 45.46  & 93.79 & -20.72 & 96.71 & 39.41  & -97.33  & 31.61 & 7 & -30.84 \\ 
& $I_{\rm ZL}$ &  &  &  21.704 $\pm$ 0.037 &  24.933 $\pm$ 0.037 & 7.560 $\pm$ 0.258 & &  &  &  &  & & &  &   &  &  &  & \\
& $I_{\rm DGL}$  &  &  &  21.805 $\pm$ 0.046 &  25.035 $\pm$ 0.046 & 6.885 $\pm$ 0.291 & &  &  &  &  & & & &  &   &  &  &  \\
& $I_{\rm A}$ to zenith &  &  &  22.280 $\pm$ 0.046 &  25.509 $\pm$ 0.046 &  4.446 $\pm$ 0.188  &  &  &  &  & & &  & &   & &  &  & \\  
 2016-07-01, 12:47:37 & $I_{\rm bkg, raw}$ & Red & 640.0  & 21.084 $\pm$ 0.002 & 24.313 $\pm$ 0.002 & 13.378 $\pm$ 0.025 & 4.282 $\pm$ 0.428  &  1.180 $\pm$ 0.236 & 34.40  & 42.84 & 139.33 & 17.56 & -1.92  & 144.32  & 8.71  & -100.08 & 31.40  & 3 & -33.87 \\
& $I_{\rm ZL}$ &  &  &  21.503 $\pm$ 0.051 & 24.732 $\pm$ 0.051 & 9.096 $\pm$ 0.429  &  & &  &  &  & & &  &   & &  & & \\
& $I_{\rm DGL}$   &  &  &  21.654 $\pm$ 0.067 & 24.883 $\pm$ 0.067 & 7.916 $\pm$ 0.490  &  & &  &  &  & & & &  &   & &  & \\
& $I_{\rm A}$ to zenith  &  &  &  22.066 $\pm$ 0.067 & 25.295 $\pm$ 0.067 &  5.416 $\pm$ 0.335  &  &  &  & & & & & &  & &  &  & \\ \hline
2019-09-18, 05:22:46 & $I_{\rm bkg, raw}$ & IR  & 811.9 & 18.895 $\pm$ 0.001 & 22.326 $\pm$ 0.001 & 100.462 $\pm$ 0.093 & 2.347 $\pm$ 0.235 &  0.382 $\pm$ 0.095 & 33.29 & 23.58 & 100.52 & 97.68 & 38.12  & 61.98 & 81.62 & -82.33  & 46.95 & 10 & -18.59\\ 
& $I_{\rm ZL}$  &  &  &  18.921 $\pm$ 0.003 & 22.351 $\pm$ 0.003 & 98.114 $\pm$ 0.252  &  &  &  &  &  & & &  &   &  &  & & \\
& $I_{\rm DGL}$    &  &  &  18.925 $\pm$ 0.003 & 22.356 $\pm$ 0.003 & 97.733 $\pm$ 0.270  &  &  &  &  &  &  & & &  &   &  &  &  \\
& $I_{\rm A}$ to zenith &  &  &  19.891 $\pm$ 0.003 & 23.322 $\pm$ 0.003 & 40.145 $\pm$ 0.111  &  &  &  &  & & &  & &   &  & & &  \\  
2016-07-01, 12:44:20 & $I_{\rm bkg, raw}$ & IR &  809.7 & 20.018 $\pm$ 0.003 & 23.247 $\pm$ 0.003 & 35.711 $\pm$ 0.099 & 4.847 $\pm$ 0.485 &   1.336 $\pm$ 0.267 & 34.48 & 42.66 & 139.29 & 17.56 & -1.92 & 144.32 & 8.71 & -100.02 & 31.41 & 3 & -33.68\\
& $I_{\rm ZL}$  &  &  &  20.176 $\pm$ 0.017 & 23.406 $\pm$ 0.017 & 30.864 $\pm$ 0.495  & & &  &  &  & & &  &   &  &  & & \\
& $I_{\rm DGL}$  &  &  &  20.224 $\pm$ 0.021 & 23.454 $\pm$ 0.021 & 29.528 $\pm$ 0.562  & & &  &  &  & & & &  &   &  &  &  \\
& $I_{\rm A}$ to zenith &  &  & 20.640 $\pm$ 0.021 & 23.870 $\pm$ 0.021 & 20.135 $\pm$ 0.382 &  &  &  &  & & &  & &   &  & &  &  \\  
\enddata
\tablenotetext{a}{Pivot wavelength of the band (nm).}
\tablenotetext{b}{Sky background in AB magnitude per arcsec$^{2}$.}
\tablenotetext{c}{Sky background in AB magnitude per pixel.}
\tablenotetext{d}{Sky background flux density in $\mu$Jansky per arcsec$^{2}$.}
\tablenotetext{e}{Zodiacal light (ZL) flux density in $\mu$Jansky per arcsec$^{2}$.}
\tablenotetext{f}{The diffuse Galactic light (DGL) in $\mu$Jansky per arcsec$^{2}$..}
\tablenotetext{g}{Gondola altitude above sea level at observation time.}
\tablenotetext{h}{Telescope elevation.}
\tablenotetext{i}{Target-moon angular separation.}
\tablenotetext{j}{Galactic longitude of target.}
\tablenotetext{k}{Galactic latitude of target.}
\tablenotetext{l}{Difference between the ecliptic longitude of the target and the ecliptic longitude of the Sun.}
\tablenotetext{m}{Ecliptic latitude of target (not to be confused with the dimensionless SED renormalization factor $\beta$).}
\tablenotetext{n}{Gondola longitude.}
\tablenotetext{o}{Gondola latitude.}
\tablenotetext{p}{Number of nights away from New Moon.}
\tablenotetext{q}{Sun altitude.}
\tablenotetext{r}{Raw measured sky background.}
\tablenotetext{s}{Sky background level with the $I_{\rm ZL}$ subtracted at the ecliptic coordinates of the target. The zodiacal light subtraction was done using data from \citet{KWON200491}.}
\tablenotetext{t}{Sky background level with the $I_{\rm ZL}$ and $I_{\rm DGL}$ subtracted at the ecliptic and Galactic coordinates of the target. The subtraction of $I_{\rm DGL}$ was done using data from \citet{toller_1981}.}
\tablenotetext{u}{Sky background light with the $I_{\rm ZL}$ and $I_{\rm DGL}$ subtracted as well as the airglow emission projected to zenith using the \textit{van Rhijn function} (see equation \ref{vanRhijn_func}).}
\end{deluxetable*}
\end{longrotatetable}

\begin{deluxetable*}{|c|c|c|c|c|c|c|c|c|}
\tabletypesize{\scriptsize}
\tablecolumns{11}
\setlength\tabcolsep{1.pt}
\label{tab:compare_all_table}
\tablecaption{Optical sky brightness measurements from mountain-top ground-based observatories (taken at zenith on moonless nights at high Galactic and high ecliptic latitudes) and from the stratosphere measured by the SuperBIT balloon-borne telescope. The magnitudes are in units of mag arcsec$^{-2}$. The stratospheric backgrounds have been zodiacal light subtracted at the ecliptic coordinates and the diffuse Galactic light subtracted at the Galactic coordinates, and the airglow has been projected to zenith using the van Rhijn function. The stratospheric brightness was measured around 5.5 hours, 3 hours, and 2 hours before the local sunrise time in 2016, 2018, and 2019 respectively. The average solar altitude angle during observations was -34$^{\circ}$, -30$^{\circ}$, and -19$^{\circ}$ above the horizon in 2016, 2018, and 2019 respectively.}
\tablehead{
\colhead{Observatory} & \colhead{Alt.} & \colhead{$N_{\rm Moon}$\tablenotemark{\scriptsize a}} & \colhead{$U$} & \colhead{$B$} & \colhead{$V$} & \colhead{$R$} & \colhead{$I$} & \colhead{Reference} \\[-0.2cm]
\colhead{} & \colhead{(km)} & \colhead{}  & \colhead{(359.7 nm)} & \colhead{(437.7 nm)} & \colhead{(548.8 nm)} & \colhead{(651.5 nm)} & \colhead{(798.1 nm)} & \colhead{} 
}
\startdata
Calar Alto & 2.2 & 0 & 22.2 & 22.6 & 21.5 & 20.6 & 18.7 & \citet{leinert_1995}\\
La Palma & 2.3 & 0 & 22.0 & 22.7 & 21.9 & 21.0 & 20.0 & \citet{Benn_1998}\\
La Silla & 2.4 & 0 & \nodata & 22.8 & 21.7 & 20.8 & 19.5 & \citet{mattila}\\
Paranal & 2.6 & 0 & 22.3 & 22.6 & 21.6 & 20.9 & 19.7 & \citet{patat}\\
Mauna Kea & 2.8 & 0 & \nodata & 22.5 & 21.6 & \nodata & \nodata & \citet{Krisciunas_1997}\\
Dome A & 4.1 & 0 & \nodata & 22.5 & 21.4 & 20.1 & \nodata & \citet{Yang_2017}\\
SuperBIT (2018)\tablenotemark{\scriptsize b} & 28.67 & 7 & \nodata & 24.078 $\pm$ 0.140 & 22.855 $\pm$ 0.067 & 22.280 $\pm$ 0.046 & \nodata & \nodata\\
SuperBIT (2019) & 33.53 & 10 & 22.174 $\pm$ 0.012 & 23.604 $\pm$ 0.041 & 22.490 $\pm$ 0.022 & \nodata & 19.891 $\pm$ 0.003 & \nodata\\
SuperBIT (2016) & 34.33 & 3 & \nodata & 25.540 $\pm$ 0.884 & 22.944 $\pm$ 0.122 & 22.066 $\pm$ 0.067 & 20.640 $\pm$ 0.021 & \nodata\\
\enddata
\tablenotetext{a}{The number of nights away from New Moon the observations were taken for the sky brightness estimates.}
\tablenotetext{b}{Note that the SuperBIT pivot wavelengths are slightly different than the Johnson-Cousins \textit{UBVRI} pivot wavelengths. See Table \ref{tab:bandpass_wavelengths} for details. Here, we present SuperBIT's UV, Blue, Green, Red, and IR results under \textit{UBVRI} for simplicity.}
\end{deluxetable*}

\begin{figure*}[htb!]
    \centering
    \includegraphics[width=45em]{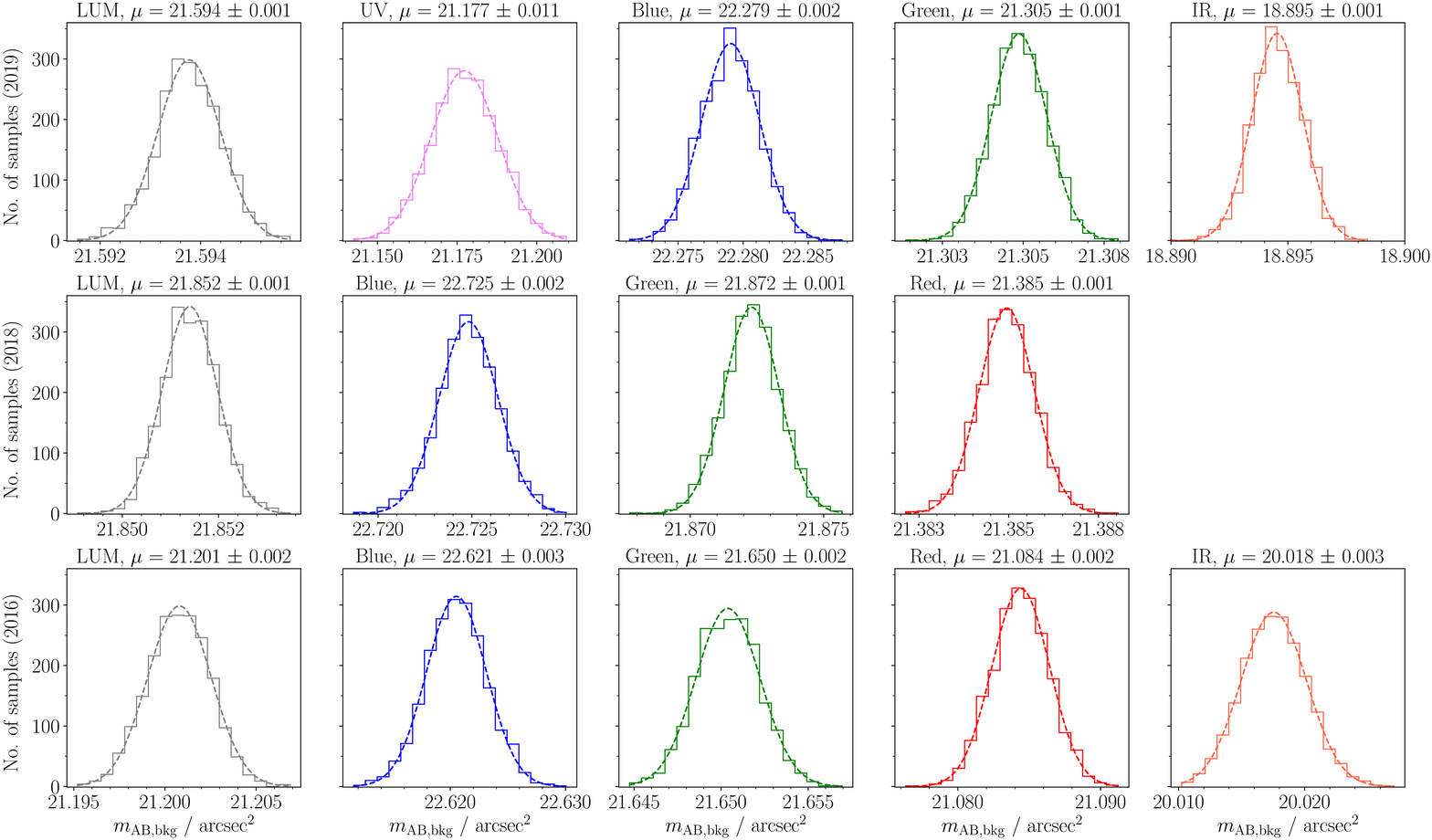}
    \caption{The raw measured sky background level in AB mag per arcsec$^{2}$ taken from the stratosphere for different bands of \superbit{} from three commissioning flights in 2016, 2018, and 2019. The histograms are the result of 2000 Monte Carlo simulations.}
    \label{fig:mc_bkg_asec}
\end{figure*}

Table \ref{tab:calibration_all} shows the calibration parameters for different bands and years. Figure \ref{fig:mc_bkg_asec} shows the raw observed sky background level from the MC simulations for different bands and years. Table \ref{bkg_all_years} shows the sky background level along with the gondola altitude, telescope elevation, moon-target angular separation, Galactic and ecliptic coordinates of the target, geographic coordinates of the gondola, and the number of nights away from New Moon the observation was taken, and the solar altitude angle. 

To ensure that the comparison between the stratospheric brightness measurements and the ground-based measurements (taken at zenith on moonless nights and at high Galactic and high ecliptic latitudes) is valid, Table \ref{bkg_all_years} presents

\begin{enumerate}
    \item \textit{Raw observed sky background}
    \item \textit{Zodiacal light subtracted sky background}
    \item \textit{Zodiacal light, diffuse Galactic light subtracted sky background}
    \item \textit{Zodiacal light, diffuse Galactic light subtracted sky background with the airglow projected to zenith}
\end{enumerate}

The zodiacal light brightness subtraction was done using observed optical zodiacal light brightness measurements from Table 3 in \citet{KWON200491} at the ecliptic longitude ($\lambda - \lambda_{\odot}$) and ecliptic latitude ($\beta$) of the target. The zodiacal brightness values in Table 3 in \citet{KWON200491} are provided in units of $S_{10}$(V)$_{\rm G2V}$. The $S_{10}$(V)$_{\rm G2V}$ unit represents the brightness equivalent to the flux of a solar type (G2V) star of tenth magnitude per square degree at the mean solar distance \citep{sparrow_unit}, and $V$ refers to the visual color in the \textit{UBV} system defined by \citet{johnson_morgan_1953}. To convert the $S_{10}$(V)$_{\rm G2V}$ units from Table 3 in \citet{KWON200491} to units of W m$^{-2}$ sr$^{-1}$ $\mu$m (and subsequently to units of $\mu$Jy arcsec$^{-2}$), we used the $S_{10}$(V)$_{\rm G2V}$ conversion factors provided as a function of wavelength in Table 2 in \citet{leinert}. 

\begin{figure*}[htb!]
    \centering
    \includegraphics[width=45em]{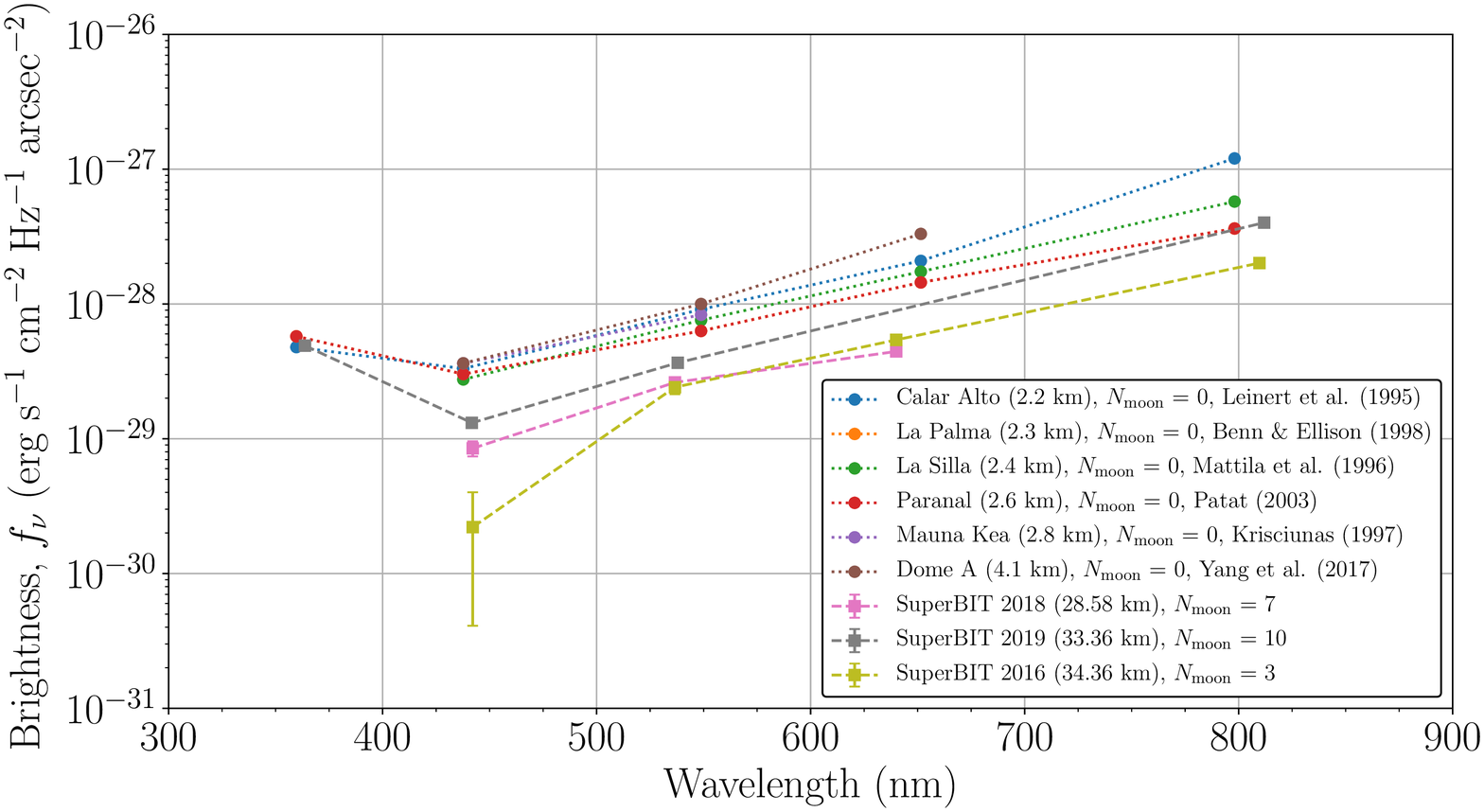}
    \caption{Optical sky brightness levels measured from mountain-top ground-based observatories and from the stratosphere using observations from the \superbit{} balloon-borne telescope. The ground-based brightness levels are based on measurements taken on the darkest, moonless nights at zenith and at high Galactic and ecliptic latitudes. To ensure that the comparison between the stratospheric brightness measurements and the ground-based measurements is valid, the stratospheric brightness levels presented here were zodiacal light and diffuse Galactic light subtracted, and the airglow was projected to zenith using the \textit{van Rhijn} function. $N_{\rm moon}$ is the number of nights away from New Moon on the night the \superbit{} observations were taken.  The stratospheric brightness was measured around 5.5 hours, 3 hours, and 2 hours before the local sunrise time in 2016, 2018, and 2019 respectively. The average solar altitude angle during observations was -34$^{\circ}$, -30$^{\circ}$, and -19$^{\circ}$ above the horizon in 2016, 2018, and 2019 respectively.  The brightness flux density presented is in units of erg s$^{-1}$ cm$^{-2}$ Hz$^{-1}$ arcsec$^{-2}$, which can be converted to AB magnitude per arcsec$^{2}$ using equation \ref{AB_mag_eqn}.}
    \label{fig:overall}
\end{figure*}

The diffuse Galactic light subtraction was done using estimates of $I_{\rm DGL}$ as a function of Galactic latitude given in Figure 76 in \citet{leinert}, which is based on Pioneer 10 measurements (see \textsection{}\ref{sec:intro} for further details). The $I_{\rm DGL}$  intensities in Figure 76 in \citet{leinert} are also given in $S_{10}$(V)$_{\rm G2V}$ units, and the conversion to W m$^{-2}$ sr$^{-1}$ $\mu$m was also done using Table 2 in \citet{leinert}. Finally, the projection of airglow to zenith was done using the \textit{van Rhijn} function (see equation \ref{vanRhijn_func}). We found $\pm3 \sigma$ clipping to be effective at removing the brightness contribution from \textit{resolved} stars, but we did not correct for $I_{\rm ISL}$ because separating the contribution of \textit{unresolved} stars is difficult. Table \ref{tab:compare_all_table} lists the sky brightness measurements from mountain-top ground (on the darkest moonless nights taken at zenith and high Galactic and high ecliptic latitudes) as well as the stratospheric brightness levels (with the subtraction of $I_{\rm ZL}$ and $I_{\rm DGL}$, and $I_{\rm A}$ projected to zenith). Figure \ref{fig:overall} compares the sky background levels measured from mountain top ground-based observatories and the stratosphere.

The stratospheric brightness was measured around 5.5 hours, 3 hours, and 2 hours before the local sunrise time in 2016, 2018, and 2019 respectively. The average solar altitude angle during observations was -34$^{\circ}$, -30$^{\circ}$, and -19$^{\circ}$ in 2016, 2018, and 2019 respectively. The $B$, $V$, $R$, and $I$ brightness levels in 2016 were 2.7, 1.0, 1.1, and 0.6 mag arcsec$^{-2}$ darker than the darkest ground-based measurements. The $B$, $V$, and $R$ brightness levels in 2018 were 1.3, 1.0, and 1.3 mag arcsec$^{-2}$ darker than the darkest ground-based measurements.  The $U$ and $I$ brightness levels in 2019 were 0.1 mag arcsec$^{-2}$ brighter than the darkest ground-based measurements, whereas the $B$ and $V$ brightness levels were 0.8 and 0.6 mag arcsec$^{-2}$ darker than the darkest ground-based measurements. The stratospheric results are consistent with the near-IR sky being generally brighter than the other optical bands because it is dominated by emission lines induced by OH and O$_{2}$ molecules \citep{meinel_one, meinel_two, Moreels2008, sul_IR, Oliva_2015}. 

\begin{figure*} [htb!]
\centering
{ \includegraphics[width=0.8\textwidth]{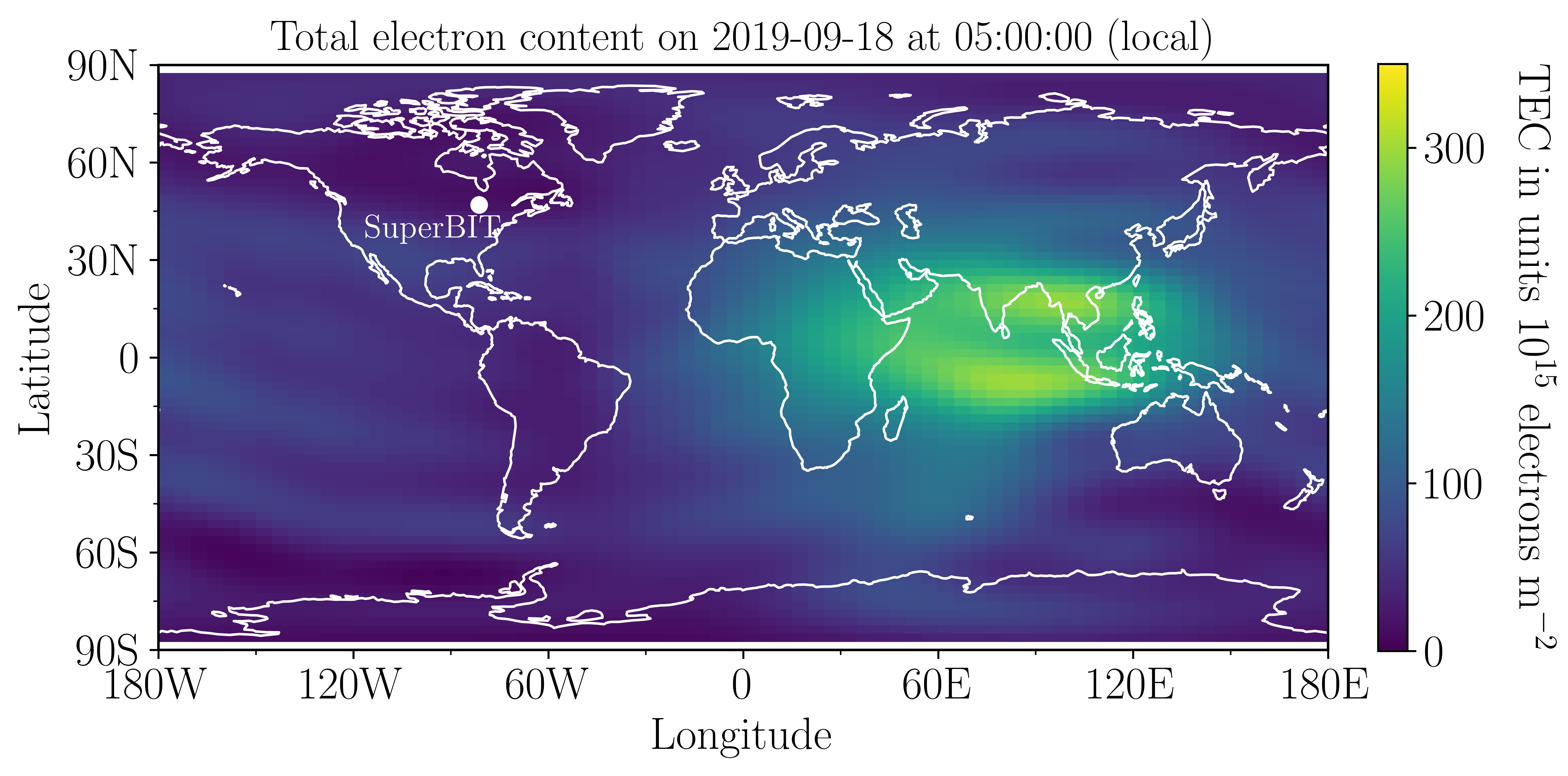}%
  }\par\medskip
{  \includegraphics[width=0.8\textwidth]{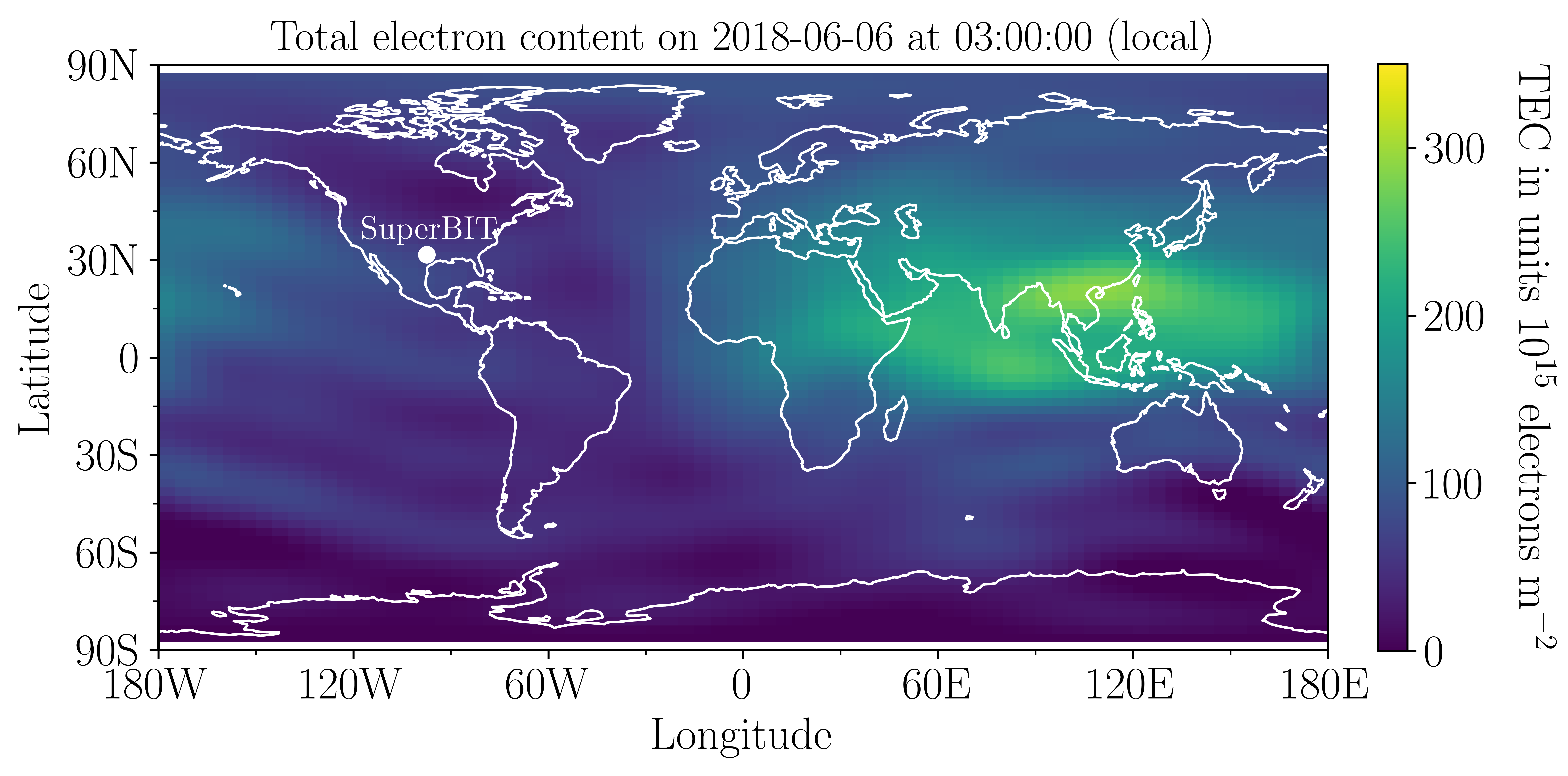}%
  }\par\medskip        
{  \includegraphics[width=0.8\textwidth]{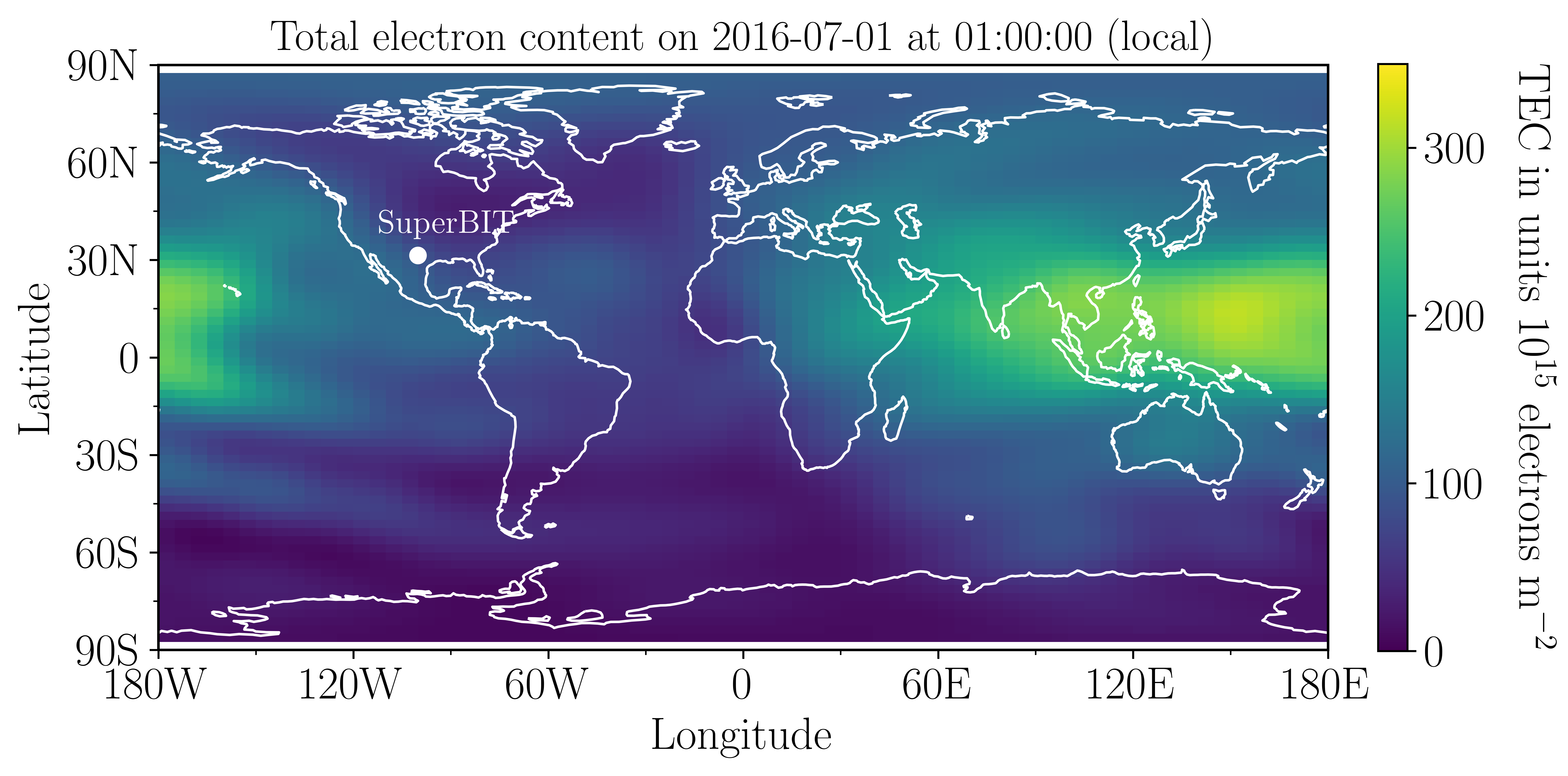}%
  }
\caption{The global total electron content (TEC) in the ionosphere at approximately the one-hour window during which the observations to estimate the sky brightness were taken during the 2016, 2018, and 2019 flights. The TEC data is taken from the International Global Navigation Satellite Systems (GNSS) service (IGS) using the rapid high-rate solution at a cadence of one map per hour provided the European Space Agency data analysis center. The geographical location of the \superbit{} gondola is shown as the white dot. During the night time observations for the three years, the TEC was roughly the same over the three years and was low compared to equatorial regions in Asia where the Sun was above the horizon.}
\label{fig:TEC}
\end{figure*}

To investigate the affect of airglow on the sky brightness, we considered the total electron density in the ionosphere during the observations. Higher ionospheric electron densities could lead to an increased probability of radiative recombination-driven lines such as oxygen and sodium lines, further increasing the sky brightness. Figure \ref{fig:TEC} shows the global total electron content (TEC) in the ionosphere at approximately the one-hour window during which the observations to estimate the sky brightness were taken during the 2016, 2018, and 2019 flights. The TEC data was taken from the International Global Navigation Satellite Systems (GNSS) service (IGS) using the rapid high-rate solution at a cadence of one map per hour provided the European Space Agency data analysis center \citep{tec}. During the \superbit{} observations taken during the night for the three years, Figure \ref{fig:TEC} shows that the TEC was roughly comparable and relatively low compared to the equatorial regions in Asia where at the time the Sun would have been above the horizon.

\section{Summary}
This paper presents the optical night sky background measurements from stratospheric altitudes with CCD images taken with the \superbit{} balloon-borne telescope. The backgrounds were estimated using data from three different commissioning flights in 2016, 2018, and 2019 at altitudes ranging from 28 to 34 km above sea level. The ground-based brightness levels are based on measurements taken on the darkest, moonless nights at zenith and at high Galactic and ecliptic latitudes. To ensure that the comparison between the stratospheric brightness measurements and the ground-based measurements is valid, the stratospheric brightness levels were zodiacal light and diffuse Galactic light subtracted, and the airglow was projected to zenith using the \textit{van Rhijn} function. The stratospheric brightness was measured around 5.5 hours, 3 hours, and 2 hours before the local sunrise time in 2016, 2018, and 2019 respectively. The average solar altitude angle during observations was -34$^{\circ}$, -30$^{\circ}$, and -19$^{\circ}$ in 2016, 2018, and 2019 respectively.

The $B$, $V$, $R$, and $I$ brightness levels in 2016 were 2.7, 1.0, 1.1, and 0.6 mag arcsec$^{-2}$ darker than the darkest ground-based measurements. The $B$, $V$, and $R$ brightness levels in 2018 were 1.3, 1.0, and 1.3 mag arcsec$^{-2}$ darker than the darkest ground-based measurements.  The $U$ and $I$ brightness levels in 2019 were 0.1 mag arcsec$^{-2}$ brighter than the darkest ground-based measurements, whereas the $B$ and $V$ brightness levels were 0.8 and 0.6 mag arcsec$^{-2}$ darker than the darkest ground-based measurements. 

The lower sky brightness backgrounds, stable photometry, and lower atmospheric absorption make stratospheric observations from a balloon-borne platform a unique tool for astronomy. This work will be continued in a future mid-latitude long duration balloon flight with SuperBIT. We plan to survey a sample of nearly 100 clusters using weak- and strong-lensing to determine their masses.  This uniform catalog will enable a qualitatively new understanding of a variety of cluster mass-observable relationships, which play a crucial role in cluster cosmology. SuperBIT observations of galaxy clusters also have the potential of improving our understanding of the nature of dark matter.

\acknowledgments
Support for the development of \superbit{} is provided by NASA through APRA grant NNX16AF65G. 
Launch and operational support for the sequence of test flights from Palestine, Texas are provided by the Columbia Scientific Balloon Facility (CSBF) under contract from NASA's Balloon Program Office (BPO).
Launch and operational support for test flights from Timmins, Ontario are provided by the \textit{Centre National d'\'Etudes Spatiales} (CNES) and the \textit{Canadian Space Agency} (CSA).

JR, EH, and JM are supported by JPL, which is run under a contract by Caltech for NASA. Canadian coauthors acknowledge support from the Canadian Institute for Advanced Research (CIFAR) as well as the Natural Science and Engineering Research Council (NSERC). LJR is supported by the Natural Science and Engineering Research Council Post-doctoral Fellowship [NSERC PDF--532579--2019]. The Dunlap Institute is funded through an endowment established by the David Dunlap family and the University of Toronto. UK coauthors acknowledge funding from the Durham University Astronomy Projects Award, the Van Mildert College Trust, STFC [grant ST/P000541/1], and the Royal Society [grants UF150687 and RGF/EA/180026]. MJ is supported by the United Kingdom Research and Innovation (UKRI) Future Leaders Fellowship `Using Cosmic Beasts to uncover the Nature of Dark Matter' [grant MR/S017216/1].

This work has made use of data from the European Space Agency (ESA) mission {\it Gaia} (\url{https://www.cosmos.esa.int/gaia}), processed by the {\it Gaia} Data Processing and Analysis Consortium (DPAC, \url{https://www.cosmos.esa.int/web/gaia/dpac/consortium}). 
Funding for the DPAC has been provided by national institutions, in particular the institutions participating in the {\it Gaia} Multilateral Agreement.
Additionally, this work made use of the \texttt{SAOImage DS9} imaging application \citep{ds9}, \texttt{Astrometry.net} \citep{astrometry}, and \texttt{SExtractor} \citep{sextractor}.

\bibliographystyle{aasjournal}

\bibliography{references}

\end{document}